\documentclass[letterpaper,twocolumn,10pt]{article}
\usepackage{usenix2019_v3}

\usepackage{tikz}
\usepackage{amsmath}

\usepackage{amsfonts}
\usepackage{booktabs}
\usepackage{url}
\usepackage{placeins}
\usepackage{here}
\usepackage[labelfont=bf]{caption}
\usepackage{subcaption}
\usepackage[noabbrev,capitalize]{cleveref}
\usepackage{multirow}

\usepackage{filecontents}

\begin{document}

\date{}

\title{Virtual Secure Platform: A Five-Stage Pipeline Processor over TFHE}

\author{

{\rm Kotaro Matsuoka$^*$, Ryotaro Banno$^\dagger$,
Naoki Matsumoto$^\dagger$, Takashi Sato$^\ddagger$, Song Bian$^\ddagger$}\\\\

{\rm$^*$Undergraduate School of Electrical and Electronic Engineering, Kyoto University},\\
{\rm$^\dagger$Undergraduate School of Informatics and Mathematical Science, Kyoto University},\\
{\rm$^\ddagger$Department of Communications and Computer Engineering, Kyoto University}\\

{\rm$^*$matsuoka.kotaro@gmail.com, $^\dagger$\{ryotaro.banno, m.naoki9911\}@gmail.com},\\
{\rm$^\ddagger$\{takashi, sbian\}@easter.kyoto-u.ac.jp}
}

\maketitle

\begin{abstract}
  
We present Virtual Secure Platform (VSP),
the first comprehensive platform that implements a multi-opcode general-purpose sequential 
processor over Fully Homomorphic Encryption (FHE) for Secure Multi-Party Computation (SMPC). 
VSP protects both the data and functions on which the data are evaluated 
from the adversary in a secure computation offloading situation like cloud computing.
We proposed a complete processor architecture with a five-stage pipeline, which improves the performance of the VSP by providing more parallelism in circuit evaluation. In addition,
we also designed a custom Instruction Set Architecture (ISA)
to reduce the gate count of our processor, along with an entire set of 
toolchains to ensure that arbitrary C programs can be compiled into our 
custom ISA.
In order to speed up instruction evaluation over VSP, CMUX Memory
based ROM and RAM constructions over FHE are also proposed.
Our experiments show that both the pipelined architecture and the CMUX Memory technique
are effective in improving the performance of the proposed processor.
We provide an open-source
implementation of VSP which achieves a per-instruction latency of 
less than 1 second.
We demonstrate that compared to the best existing processor over FHE, 
our implementation runs nearly 1,600$\times$ faster. %

\end{abstract}

\section{Introduction}
\label{sec:intro}
In a typical cloud computing scheme, clients want to offload their computations, that are, the evaluations of some programs over their private data, to some cloud server. The problem we try to tackle in this paper is to protect the programs and data of the clients against the server per se, or some third-party intruder who has physical access to the server. 
Since current mainstream physical processors, like Intel Xeon, cannot directly run encrypted instructions (i.e., the program to be offloaded), encrypted functions and data must be decrypted at run-time. 
Therefore, current cloud computing schemes suffer from side channel attacks~\cite{cacheout,sgaxe}.
In addition, processor vendors may also plant backdoors. As a result,
the cloud service vendors and those who can physically access the servers are, in theory, able to steal the program along with the input data from the clients.

The key idea to solve the problem mentioned above is to directly run encrypted instructions~\cite{7725560}.
In other words, the client of the cloud service sends the encrypted instructions which represent the function to be evaluated and the encrypted inputs to the cloud sever. Meanwhile, the cloud server evaluates the function using the inputs, without decryption. After the evaluation, the cloud server sends back the encrypted results to the client. During the entire evaluation process, the cloud server does not have access to any plaintext, so the evaluated function and the data are protected.
The above scheme can be established by representing the processor as Boolean circuits, and the evaluation of the circuits are conducted through the use of Secure Multi-Party Computation (SMPC) protocols. Because Boolean circuits can be represented by a graph containing different types of logic gates as graph nodes (e.g., in~\cref{fig:graph}),
if we can perform the logical operations over encrypted input bits, we can emulate the operation of a processor by evaluating the processor circuit with the associated encrypted inputs.

Currently, we have two well-known SMPC candidates for evaluating Boolean circuits directly over encrypted inputs, namely Garbled Circuit (GC)~\cite{4568207} and Fully Homomorphic Encryption (FHE)~\cite{10.5555/1834954}.
GC implements SMPC operations by providing a set of 
encrypted truth tables for the outputs of the corresponding 
logic gates. During GC evaluation, the truth tables are
evaluated obliviously to carry out the encrypted function evaluation.
On the other hand, FHE is intrinsically more of a Secure Computation Offloading (SCO)
scheme, where inputs to some public function are encrypted. 
The evaluator directly evaluates the public function over the ciphertexts, and returns the results
to the encryption party.
There are two previous works which propose to run encrypted 
instructions using GC: 
TinyGarble~\cite{7163039} and GarbledCPU~\cite{7544316}.
These works implement a processor
with the MIPS~\cite{10.1145/1014194.800930} Instruction Set Architecture (ISA). 
Since most modern compilers support MIPS, both TinyGarble and
GarbledCPU support the evaluation of most conventional programs, 
e.g., programs written in the C language.
However, in theory, we cannot achieve SCO 
with GC, as the generation of the GC truth tables always take more 
computational resources than locally evaluating the program.
In contrast, as mentioned, FHE is inherently an SCO 
scheme~\cite{10.5555/1834954}. Unlike GC, there is no need for tables generation for the evaluation of logic gates in FHE.
To the best of our knowledge, FURISC~\cite{Chatterjee2019} is the 
only previous work which implements a processor over the Smart-Vercauteren FHE Cryptosystem~\cite{10.1007/978-3-642-13013-7_25,libScarab}.
The processor only accepts one Turing-complete instruction, Subtract Branch if Negative (SBN).
This means that it is necessary to modify modern compilers like Clang or GCC to work with FURISC,
which is a highly non-trivial task.
In fact, FURISC does not have any compiler support.

We propose Virtual Secure Platform (VSP), a comprehensive 
platform that provides a full set of tools for a complete two-party SCO 
scheme. Our standalone platform includes open-sourced designs and 
implementations of HE libraries, processor architectures, custom ISA and 
compiler environments. Building upon the well-known Torus Fully Homomorphic Encryption (TFHE) scheme, 
VSP allows any user with an arbitrary C 
program to execute their codes in an SCO manner.
To the best of our knowledge, VSP is the fastest and most 
complete (in terms of the set of tool sets we provide) open-source 
processor platform to date.

{\bf Contributions}: In brief, our contributions are as follows:
\begin{itemize}
    \item We present VSP, the first comprehensive platform that implements a multi-opcode general-purpose sequential processor over TFHE, which enables two-party SCO.
    We also provide an open-source Proof of Concept (PoC) implementation of VSP,
    including our pipelined processor.
    \item We implemented the entire toolchain including a C compiler based on LLVM
    in order to fully support C language in VSP. %
    The toolchain is based on our custom ISA named CAHPv3.
    \item We propose CMUX Memory, an optimized memory structure over TFHE. We
    fully leverage the Leveled Homomorphic Encryption (LHE) mode of the TFHE to ensure fast memory access,
    which is one of the main performance bottlenecks of VSP.
    \item Our open-source PoC implementation can evaluate one clock cycle of the 
    processor in less than 1 second.
    This translates to nearly 1,600$\times$ per-instruction latency reduction compared to FURISC, 
    the state-of-the-art FHE-based SCO scheme.
\end{itemize}

\section{Preliminaries}

In this section, we define and explain some basic concepts used throughout 
this work. We first review the properties and constructions of HE in 
\cref{sec:prelim_he}. Then, we give an overview on the security
properties of the SMPC protocols focused in this work in \cref{sec:prelim_sec}.
Finally, we briefly summarize the general terminologies involved in processor
designs in \cref{sec:prelim_proc}.

\subsection{Homomorphic Encryption}\label{sec:prelim_he}

\subsubsection{Overview of Homomorphic Encryption}
Homomorphic Encryption (HE) is a form of encryption which permits encrypted data to be evaluated without decryption~\cite{Rivest1978b}. 
HE can be classified into several categories depending on the types of functions that are permitted to be evaluated.
A Fully Homomorphic Encryption (FHE) scheme allows one to evaluate arbitrary 
functions.
Some popular FHE candidates include Torus Fully Homomorphic Encryption 
(TFHE)~\cite{Chillotti2020}, Smart-Vercauteren 
Cryptosystem~\cite{10.1007/978-3-642-13013-7_25} and Brakerski-Gentry-Vaikuntanathan 
(BGV)~\cite{10.1145/2090236.2090262}. All of the above mentioned candidates can
evaluate arbitrary Boolean circuit over encrypted ciphertexts.
Beside FHE, another category of HE is called Leveled Homomorphic Encryption (LHE). 
LHE has limitations on the depth of function that can be expressed, but are 
much faster than FHE in general.
Depth here means the number of consecutive multiplications to be performed on the same ciphertext.
Lastly, we note that some FHE schemes like TFHE and BGV have LHE modes.

In VSP, we cannot know {\it{a priori}} how many times we have to evaluate the circuit of the processor. %
This is because a general solution to the problem of determining how many clock cycles a program will take
written in a Turing-complete language solves the famous Halting Problem, which
is known to be undecidable.
Therefore, FHE is most suitable for constructing processor-like architectures
as in VSP.

{\bf Bootstrapping}:
Bootstrapping is one of the most important idea in the construction of FHE. It is 
proposed in the seminal work of Gentry~\cite{10.5555/1834954}. The bootstrapping can 
be thought as evaluating a decryption function over HE. Bootstrapping is needed
for 
FHE schemes, as we can remove the noises from the ciphertexts generated during
the evaluations. Bootstrapping needs additional keys for evaluation, 
including an encrypted secret key.

{\bf Key switching}: Key switching is a function that maps a ciphertext ${\rm Enc}_{s_1}(m)$
to ${\rm Enc}_{s_2}(m)$ without decryption, where ${\rm Enc}_{s_i}(m)$
means encrypted $m$ with a secret key $s_i$. As with bootstrapping,
this function also requires an encrypted secret key,
but its format is different from that required for bootstrapping,
because key switching needs $Enc_{s_2}(s_1)$.

In this paper, we call the set of the keys which are required to evaluate both the 
bootstrapping and the key switching as Bootstrapping Key.

\subsubsection{TFHE}
\label{sec:TFHE}

TFHE~\cite{10.1007/978-3-662-53887-6_1,Chillotti2020} is one kind of FHE. TFHE 
natively supports one-operand logic operations like NOT, two-operand logical 
operations like NAND, NOR, XNOR, AND, OR and XOR, and the three-operand MUX. 
There are two reasons for choosing TFHE as a foundation of VSP.
First, bootstrapping of TFHE only takes 10 milliseconds order. This is the fastest one to  our best knowledge. In contrast, bootstrapping of BGV takes order of minutes with HElib~\cite{cryptoeprint:2014:873}. 
Second, TFHE supports LHE mode which we find to be efficient in constructing memory units,
and a detailed construction of memory units over the LHE mode of TFHE is explained in \cref{sec:cmux}.

In what follows, we describe TFHE briefly. We will strip away unnecessary generality in order to keep the explanations straightforward.

{\bf Notations}:
In this work, we adopt a similar notation style as in~\cite{Chillotti2020}, which
is listed below.

\begin{description}
\setlength{\itemsep}{0pt}
    \item $\mathbb{B}$: The set $\{1,0\}$ without any structure.
    \item $\mathbb{T}$: The real Torus $\mathbb{R}/\mathbb{Z}$, the set of real number modulo 1. In this work,
    we define the interval of $\mathbb{T}$ to be $[-0.5,0.5)$.
    \item $\mathbb{T}_N[X]$, $\mathbb{Z}_N[X]$: The rings of polynomials $\mathbb{R}[X]/(X^N+1) \bmod 1$ and $\mathbb{Z}[X]/(X^N+1)$.
    \item \texttt{$\mathbb{B}_N[X]$}: The polynomials in $\mathbb{Z}_N[X]$ with binary coefficients.
    \item \texttt{$1[X]$}: The polynomial in $\mathbb{Z}_N[X]$ whose coefficients are all 1. 
    \item \texttt{$\mathit{sgn}(a[X])$}: The polynomial whose $i$ th coefficient is $\mathit{sgn}(i\text{ th coefficient of }a[X])$.
    \item $E^p$: The set of vectors of dimension $p$ with entries in $E$.
    \item $M_{p,q}(E)$: The set of $p \times q$-size matrices with elements in $E$.
    \item $U(E)$: The uniform distribution over $E$.
    \item $\leftarrow$: $x\leftarrow D$ means $x$ itself or its entries or coefficients are drawn from the distribution $D$.
    \item $n,N,l,\alpha,\mu$: $n,N,l\in\mathbb{N}$, $\alpha \in \mathbb{R}$ and $\mu=1/8$.
    \item $\mathbf{a},a[X],b[X]$: $\mathbf{a}\in\mathbb{T}^n$ and $a[X],b[X]\in\mathbb{T}_N[X]$.
\end{description}

{\bf Modular Gaussian Distribution}:
Let $k \geq 1$ and $\sigma \in \mathbb{R}^+$. For all $\mathbf{x}\in \mathbb{R}^k$, 
we refer to 
the Gaussian function of center 0 and standard deviation $\sigma$ as
$\rho_{\mathbb{R}^k,\sigma}(\mathbf{x}) = \exp(-\|\mathbf{x}\|^2/2\sigma^2)$. 
Meanwhile, $\mathcal{D}_{\mathbb{T}^k,\sigma}(\mathbf{x})$ defines a (restricted) Gaussian Distribution of center $\mathbf{0}$ and standard deviation $\sigma$ over $\mathbb{T}^k$, and is derived by $\mathcal{D}_{\mathbb{T}^k,\sigma}(\mathbf{x}) = \sum_{l\in\mathbb{Z}} \rho_{\mathbb{R}^k,\sigma}(\mathbf{x}+l\cdot \mathbf{1})$.

{\bf TLWE}:
TLWE is the Torus version of the learning with errors (LWE) problem~\cite{10.1145/1060590.1060603}. TLWE can be represented as $(\mathbf{a},b)$, an $n+1$ dimensional Torus vector. 
$\mathbf{s}\in \mathbb{B}^n$ is the secret key and $\mathbf{s} \leftarrow \mathbb{B}$.

{\it Encryption}: Let $e \leftarrow \mathcal{D}_{\mathbb{T},\sigma}(x)$ and $\mathbf{a}\leftarrow U(\mathbb{T}^n)$. $m\in\mathbb{B}$ is the plaintext message. Then, $b=\mathbf{a} \cdot \mathbf{s}+\mu(2m-1)+e$.

{\it Decryption}: Return $(1+\mathit{sgn}(b-\mathbf{a}\cdot\mathbf{s}))/2$.

{\bf TRLWE}:
TRLWE is the Torus version of ring-LWE. TRLWE can be represented as $(a[X],b[X])$, a two dimensional Torus polynomial vector. 
$s[X]\in \mathbb{T}_N[X]$ represents the secret key and $s[X] \leftarrow U(\mathbb{B})$.

{\it Encryption}: Let $e[X]\in\mathbb{T}_N[X] \leftarrow \mathcal{D}_{\mathbb{T}^N,\sigma}(\mathbf{x})$ and $a[X] \leftarrow U(\mathbb{T}^N)$.  $m[X]\in\mathbb{B}_N[X]$ is the plaintext message. Then, $b[X] = a[x]\cdot s[X] + \mu(2m[X]-1[X]) +e[X]$

{\it Decryption}: 
Return $(1[X]+\mathit{sgn}(b[X]-a[X]\cdot s[X]))/2$.

{\bf TRGSW}:
This is a Torus and ring version of GSW, which is represented as a vector of TRLWE ciphertexts, or equivalently, a matrix of polynomials. TRGSW ciphertexts are in $M_{2,2l}(\mathbb{T}_N[X])$. 

{\it Encryption}: Let $l,Bg \in \mathbb{N}, i \in [1,2l]$. $e[X]\in\mathbb{T}_N[X] \leftarrow \mathcal{D}_{\mathbb{T}^N,\sigma}(\mathbf{x})$ and $a[X] \leftarrow U(\mathbb{T}^N)$. $m\in\mathbb{B}$ is the plaintext message. Then, $b_i[X] = a_i[x]\cdot s[X] + e_i[X]$ and the ciphertext $\mathbf{C}$ is defined as follows: 
\[\mathbf{C} = \begin{pmatrix}
      a_1[X]+\frac{m}{Bg} & b_1[X] \\
      a_2[X]+\frac{m}{Bg^2} & b_2[X] \\
      \vdots & \vdots\\
      a_l[X] +\frac{m}{Bg^l} & b_l[X]\\
      a_{l+1}[X] & b_{l+1}[X] +\frac{m}{Bg}\\
      \vdots & \vdots\\
      a_{2l}[X] & b_{2l}[X] +\frac{m}{Bg^l}
    \end{pmatrix}\]
We omit the explanation on the decryption of TRGSW as it is not needed in this paper.

{\bf Sample Extraction and Identity Key Switching}:
This operation converts a TRLWE ciphertext into a TLWE ciphertext. Identity Key Switching  (IKS) denotes the special case of Public Key Switching where the public function is the identity function~\cite{Chillotti2020}. 
The noise variance of the output TLWE ciphertext becomes larger than the input TRLWE ciphertext because IKS adds noises.
The construction of Bootstrapping in TFHE uses this as a fundamental block.

{\bf Bootstrapping in TFHE}:
In TFHE, Bootstrapping can be defined for TRLWE and TLWE. This can be configured by Sample Extraction (SE) and IKS at the beginning or end of the Bootstrapping procedure. The type of output ciphertext is the same as that of the input but the noise is refreshed to
the level of a freshly encrypted ciphertext.

{\bf CMUX}:
CMUX is short for Controlled MUltipleXer.
This is one of the LHE-mode operations of TFHE and is equivalent to a homomorphic multiplexer.
CMUX takes two TRLWE ciphertexts as inputs and a TRGSW ciphertext as its selector input.
CMUX outputs a TRLWE cipehrtext. The noise variance of the output TRLWE ciphertext is bigger than that of the inputs
because additional noise is induced by the CMUX operation.

{\bf Homomorphic Gates}:
These are FHE-mode operations of TFHE and they represent logic gates. Their inputs and outputs are TLWE ciphertexts.
All Homomorphic Gates except for HomNOT perform bootstrapping in their evaluation.
HomNOT only negates the coefficients of its input TLWE ciphertext,
so the noise variances remain the same for its input and output ciphertexts.

{\bf HomMUX without SE and IKS}:
This is MUX of Homomorphic Gates (HomMUX) without SE and IKS in its Bootstrapping.
By definition, HomMUX without SE and IKS maps three TLWE ciphertexts to a TRLWE ciphertext.
HomMUX without SE and IKS is used in the construction of our CMUX Memory.

{\bf Circuit Bootstrapping}:
This is a function which converts a TLWE ciphertext into TRGSW ciphertext
proposed in~\cite{Chillotti2020}. %
The noise variance is always the same between the input and output
of Circuit Bootstrapping, as bootstrapping is performed during the process.

{\bf Parameters of TFHE}:
Parameters of TFHE are one of the most important things in security analysis of VSP
since they determine the security level of TFHE. In our PoC implementation,
we adopt parameters recommended in~\cite{10.1007/978-3-662-53887-6_1,TFHE}.
The estimated security of the parameter set is 
80-bit~\cite{Chillotti2020, TFHEparams}.

\subsection{Terms for Security Analysis}\label{sec:prelim_sec}

\subsubsection{Definitions for Protocols}
The main protocol we treat in this paper is two-party Secure Computation Offloading (SCO).
Two-party SCO is a special case of Private Function Evaluation (PFE)~\cite{10.1007/978-3-642-38348-9_33}. To clarify the difference between VSP and GarbledCPU or TinyGarble, we also explain Private Function Secure Function Evaluation (PF-SFE).

{\bf Definition of Alice and Bob}: In this paper, Bob is someone who provides most of computational resource, like cloud vendors, and Alice is someone who is the user of the
cloud service and possesses the secret key. Both of them are interested in learning as much private information as possible from the other party.

{\bf Two-party SCO}: In this protocol, only Alice has private information, which is a function to be evaluated along with the inputs. Furthermore, only Alice learns the result of the evaluation. 

{\bf Two-party PF-SFE}: In this protocol, Bob has a  function to be evaluated and Alice has its input, but only Alice learns the result of the function.

\subsubsection{Security Assumptions}

There are two assumptions in our security analysis:
the 1-circular security defined in~\cite{10.1007/978-3-642-30057-8_32}, which relates to security of the TFHE scheme, and
the honest-but-curious model, which limits the behavior of the adversary.

{\bf 1-circular security}: Circular security is classified into 
Key-Dependent-Message (KDM) security~\cite{10.5555/646558.694914,10.1007/978-3-642-30057-8_32}.
1-circular security means that encryption of a secret key using the secret key itself is secure. This is assumed in~\cite{Chillotti2020} to simplify the implementation.

{\bf Honest-but-curious model}: A honest-but-curious adversary is a legitimate participant in a communication protocol who will not deviate from the defined protocol but will attempt to learn all possible information from legitimately received messages~\cite{HBC}.

\subsection{Terms for Processor Design}\label{sec:prelim_proc}

\begin{figure}[t]\centering
\begin{minipage}[b]{0.45\linewidth}
  \centering
  \includegraphics[width=\linewidth]{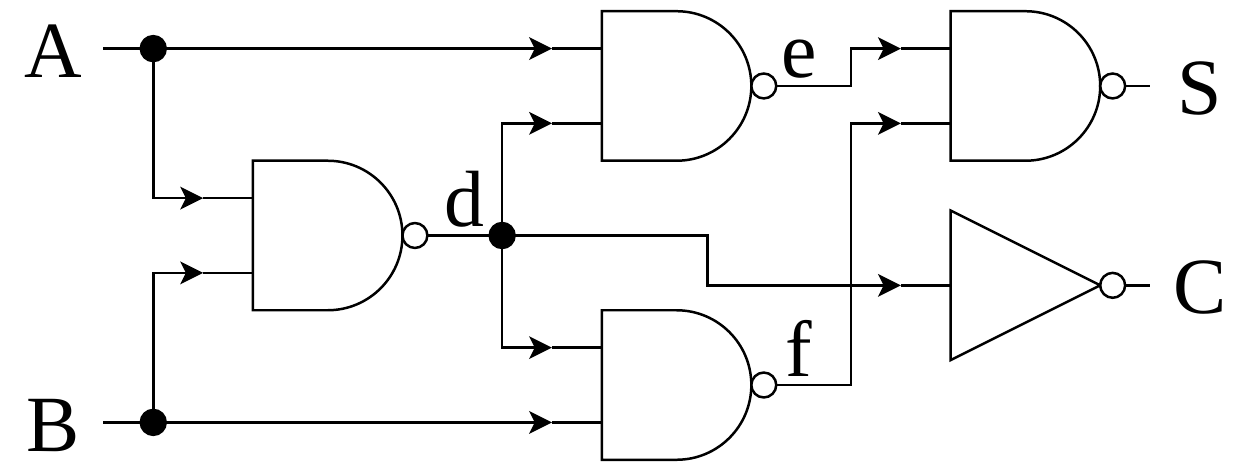}
  \subcaption{Circuit representation}
  \label{fig:halfadder}
  \end{minipage}
  \begin{minipage}[b]{0.45\linewidth}
    \centering
    \includegraphics[width=\linewidth]{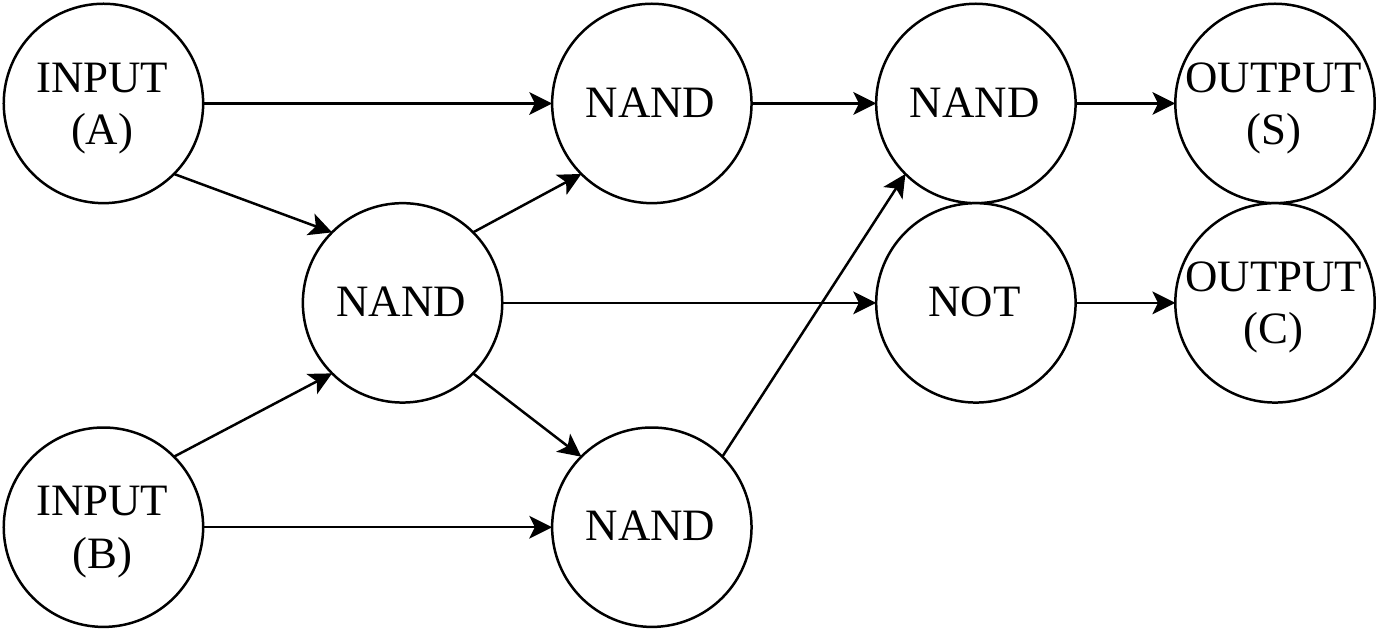}
    \subcaption{Graph representation}
    \label{fig:graph}
\end{minipage}
\caption{The (a) circuit and (b) graph representation of a half adder.}
\end{figure}

We use a half adder as an example for explaining circuit-related vocabulary in this paper.
A half-adder can be represented as \cref{fig:halfadder}.
To simplify the explanation, we only use NAND and NOT gates here.
We denote its input bits by $A$ and $B$, and its output bits by $S$ and $C$.
A half adder computes a 1-bit addition. For example, if $A=B=1$, then $S=0,C=1$, which calculates one plus one equals two.
Let $e,d,f$ denote intermediate outputs of the gates. If we represent a NAND gate by function ${\rm NAND}(\cdot,\cdot)$ and NOT gate by $\mathrm{NOT}(\cdot)$, we can interpret this circuit as a series of equations like the following:
\begin{align}
    \begin{cases}
    d &= {\rm NAND}(A,B)\\
    e &= {\rm NAND}(A,d)\\
    f &= {\rm NAND}(d,B)\\
    S &= {\rm NAND}(e,f)\\
    C &= {\rm NOT}(d)
    \end{cases}
    \label{eq:halfadder}
\end{align}

{\bf Boolean Circuits over TFHE}:
The main idea for evaluating Boolean circuits over TFHE is replacing each logic gate in the Boolean circuit by a Homomorphic Gate from TFHE.
In the half-adder circuit shown in \cref{fig:halfadder},
this means replacing ${\rm NAND}(\cdot,\cdot)$ and ${\rm NOT}(\cdot)$ in \cref{eq:halfadder} by equivalent TFHE operations, that is, ${\rm HomNAND}(\cdot,\cdot)$ and ${\rm HomNOT}(\cdot)$.
Let ${\rm Enc}(\cdot)$ denote encryption function of TFHE.
Then, we can reinterpret \cref{fig:halfadder} by using the idea as follows:
\begin{align*}
\begin{cases}
    {\rm Enc}(d) &= {\rm HomNAND}({\rm Enc}(A),{\rm Enc}(B))\\
    {\rm Enc}(e) &= {\rm HomNAND}({\rm Enc}(A),{\rm Enc}(d))\\
    {\rm Enc}(f) &= {\rm HomNAND}({\rm Enc}(d),{\rm Enc}(B))\\
    {\rm Enc}(S) &= {\rm HomNAND}({\rm Enc}(e),{\rm Enc}(f))\\
    {\rm Enc}(C) &= {\rm HomNOT}({\rm Enc}(d))
\end{cases}
\end{align*}
This interpretation enables us to evaluate single-bit addition over TFHE with encrypted inputs and outputs. We can formulate, in a similar way, an entire processor circuit over TFHE.

{\bf Pipeline}:
Pipeline is a mechanism to increase the number of gates that can be evaluated in parallel ($\mathfrak{g}$) by 
dividing the circuit into several stages with registers. The registers hold the inputs to and 
outputs from the stages for synchronization. 

\begin{figure}[t]\centering
\begin{minipage}[b]{0.45\linewidth}
  \centering
  \includegraphics[width=\linewidth]{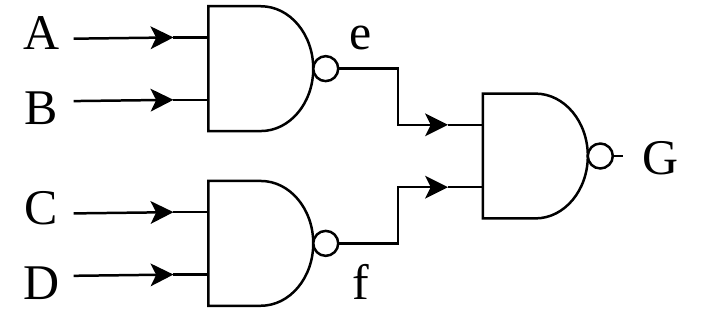}
  \subcaption{Unpipelined}
  \label{fig:normal}
  \end{minipage}
  \begin{minipage}[b]{0.45\linewidth}
    \centering
    \includegraphics[width=\linewidth]{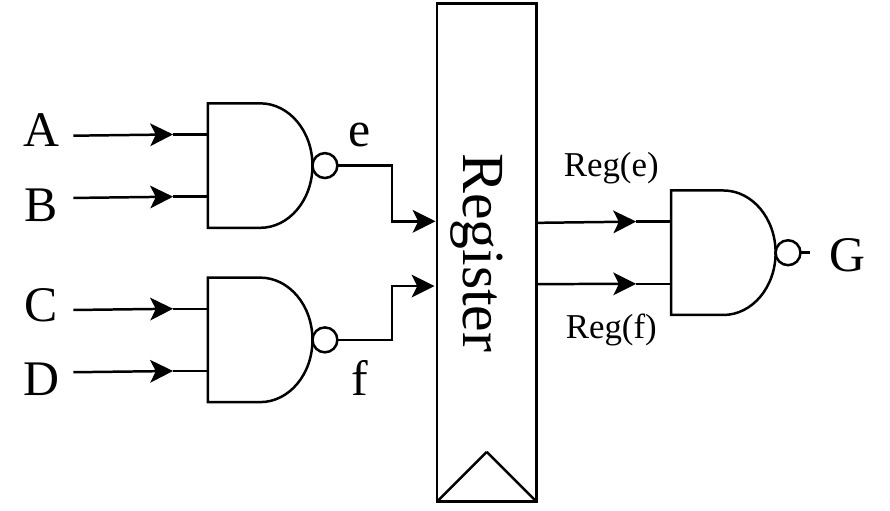}
    \subcaption{Pipelined}
    \label{fig:pipelined}
\end{minipage}
\caption{Examples of (a) unpipelined and (b) pipelined circuits.}
\label{fig:pipeline-circuit}
\end{figure}

\cref{fig:pipeline-circuit} shows the unpiplined and pipelined circuits.
In the unpipelined circuit, $\mathfrak{g} =2$ because only $\mathrm{NAND}(A, B)$ and $\mathrm{NAND}(C, D)$ can be evaluated simultaneously.
Meanwhile, in the pipelined circuit, $\mathfrak{g}=3$ because the register feeds the value to the NAND gate, such that $\mathrm{NAND}(\mathrm{Reg}(e), \mathrm{Reg}(f))$, $\mathrm{NAND}(A, B)$ and $\mathrm{NAND}(C, D)$ can be evaluated
in parallel.
That is how the pipelining increases the parallelism of the processor.
Lastly, we emphasize an important point that pipelining adds considerable costs to
physical processor designs as physical registers need to be added to the processor
circuit to enable pipelining. However, for FHE-based processors, we do not need to
implement these pipeline registers using Homomorphic Gates. The intermediate ciphertexts
can simply be stored into the physical memory, acting as a ``pipeline register.''
This reasoning holds true for all sequential elements (e.g., flip-flops) in the VSP processor architecture.

\section{Related Works}

There are some previous works which enable one to run encrypted programs by implementing a Boolean circuit of a processor over SMPC protocols.
We only provide a brief summary on the most relevant works, and more works can be found in \cref{sec:additionalrealted}.

\subsection{Processor over HE}
There have been a few works that have attempted to implement processors over HE to run encrypted instructions~\cite{6295998,7079493,DBLP:journals/ijccbs/BreuerB19,7469876}.
However, only FURISC~\cite{cryptoeprint:2015:699,Chatterjee2019} represents the processor as a Boolean circuit.
FURISC uses Smart-Vercauteren Cryptosystem~\cite{10.1007/978-3-642-13013-7_25,libScarab} to represent its processor.
Smart-Vercauteren Cryptosystem is an FHE which supports XOR and AND over the ciphertexts.
FURISC theoretically can be solutions for two-party SCO although it is not discussed in their paper~\cite{Chatterjee2019}.
FURISC implements an One Instruction Set Computer (OISC) processor which supports only one instruction, SBN.
This means modifying modern compilers like Clang or GCC to work for it is not an easy task because it is far different from current mainstream instruction sets.
In fact, there is no high-level language compiler available for FURISC.
In the experiments in \cref{sec:eval},
we show that VSP runs nearly 1,600$\times$ faster than the estimated runtime of FURISC.

\subsection{Garbled Processor}
Garbled Processor is the name for the processor over Garbled Circuit (GC).
There are three works, ARM2GC~\cite{Songhori2019ARM2GCSG}, TinyGarble~\cite{7163039}, and GarbledCPU~\cite{7544316}.
ARM2GC emulates an ARM processor, but it assumes the function to be evaluated as public.
TinyGable and GarbledCPU emulate a MIPS processor and enable to use conventional programming representation for two-party PF-SFE~\cite{7163039,7544316}. The
most critical weakness of Garbled Processors is that, in theory, 
such constructions cannot achieve two-party SCO. If Garbled Processor is used in SCO, Alice needs to generate a table of ciphertexts for all of the outputs of each gate for each clock cycle. This means Alice has to do more computationally intensive tasks than directly evaluating the function with the inputs.

\section{Abstract Protocol Flow in Two-party SCO}

In this section, we explain how VSP works in the two-party SCO protocol.
Two-party PF-SFE can be theoretically achieved by modifying two-party SCO. See \cref{sec:pf-sfe}.

{\bf{Public/Private Data}}:
The parameters of TFHE, Bootstrapping Key, the circuit of the processor,
the upper-bound of the number of processor evaluation,
the ciphertexts of ROM and RAM, and the sizes of ROM and RAM are public to all parties.
The plaintext data of ROM and RAM data, the result of the evaluated function and the secret key are private for Alice.

\subsection{Abstract Protocol Flow}\label{subsec:abstract-protocol-flow}

\begin{figure}[t]
  \centering
  \includegraphics[width=0.8\linewidth]{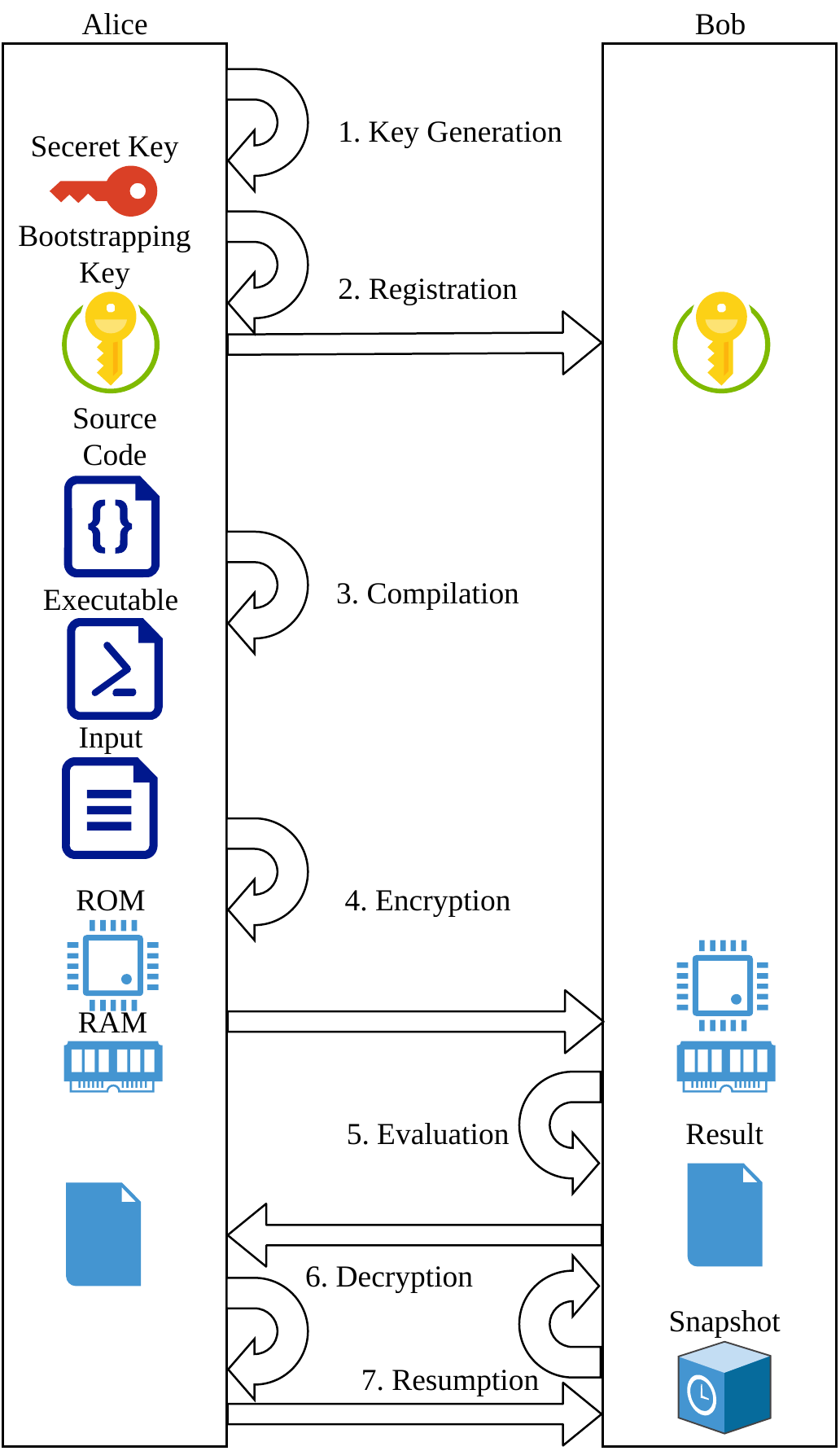}
  \caption{The proposed protocol flow of two-party SCO.}
  \label{fig:scoworkflow}
\end{figure}

The protocol flow of VSP can be divided into seven phases,
and a visual depiction is shown in \cref{fig:scoworkflow}. The phases
are discussed as follows.

\begin{enumerate}
    \item {\bf Key Generation}: Alice generates a secret key.
    \item {\bf Registration}: Alice generates a Bootstrapping Key from the secret key and  sends the Bootstrapping Key to Bob.
    \item {\bf Compilation}: Alice compiles the source code of the function to be evaluated into executable (instructions) for the processor using an ordinary compiler. 
    \item {\bf Encryption}: Alice combines the executable with inputs, and encrypts them as ROM and RAM. The executable has a RAM part because of the initialization of global variables. In this phase, Alice also decides how many clock cycles Bob has to evaluate.
    \item {\bf Evaluation}: Bob evaluates the encrypted ROM and RAM by repeatedly evaluating the processor circuit using the TFHE ciphertexts from Alice for the designated number of clock cycles. 
    In this phase, what we refer to as the snapshot is also generated. A snapshot contains all necessary information for the Resumption phase, including ciphertexts of current register values, ROM and RAM.
    \item {\bf Decryption}: Alice decrypts the encrypted result using her secret key.
    \item {\bf Resumption}: Alice checks the termination flag which is included in the result. If the flag indicates that the evaluation of the function has finished, the protocol is terminated. If not, Alice re-generates the number of clock cycles Bob needs to additionally evaluate the processor circuits. 
    Then, Bob executes the evaluation for the designated clock cycles using the information contained in the snapshot and returns to Decryption phase.
\end{enumerate}

In the above procedures, 1. and 2. are needed only once. If Alice wants to evaluate multiple sets of functions and (or) inputs, the secret key and the Bootstrapping Key can be reused. Therefore, the computational and communication costs for them are negligible.

{\bf{Client-Side Computation and Outsourcing}}: 
Here, we briefly show why VSP is able to provide a meaningful computation outsourcing scheme.
To outsource a program in a meaningful way, the cost of client-side (i.e., Alice-side) computations for setting up the outsourcing protocol must be less than that of locally evaluating the program to be outsourced. 
In VSP, the client-side costs almost entirely depend on the security parameter and the size of the memory $m$, but not on the number of clock cycles $n$ required to evaluate the compiled program. 
Therefore, for any program where $k\cdot m\leq o(n)$ for some constant $k$ ($k$ only depends on the security parameter),  it holds that 
the client-side computation costs are a less than that of directly evaluating the program.

\section{Security Analysis}
\label{sec:security}

In this section, we analyze security of VSP. We also describe the termination problem, which is one of the reasons why we assume honest-but-curious adversary model. In this paper, we also assume 1-circular security as assumed in TFHE.

\subsection{Security Analysis in Two-party SCO}

In this paper, we assume that Bob has physical access to the computational resource.
More precisely, the assumption is that Bob can read even electric signals in the CPU dies between transistors.
Therefore, any private information which is decrypted in the computational resource leaks to Bob. 

Bob tries to guess Bootstrapping Key, ROM, RAM, registers, wires, etc. However, since we assume honest-but-curious adversary model, this can be reduced to the hardness of decryption of ciphertexts of TFHE in Chosen-plaintext Attack (CPA) setting. 
As LWE-based FHE schemes are generally based on well-established hardness assumptions, 
the security of VSP can be easily guaranteed.

\subsection{The Termination Problem}
\label{sec:termprob}
In VSP, it is obvious that Bob cannot know if the evaluated program is halted or not, 
without run-time communication with Alice, as the state of the
processor is entirely encrypted.
The termination problem is also discussed in FURISC paper~\cite{Chatterjee2019}.
The protocol which is claimed to be a solution for the problem in the paper can be interpreted as the following procedures in VSP:

\begin{enumerate}
    \item Bob sends to Alice a TLWE ciphertext of the termination flag.
    Here, the termination flag indicates if the function evaluation is 
    finished or not.
    \item Alice decrypts the termination flag and tells Bob to terminate or continue the evaluation.
    \item If Alice decided to terminate the evaluation in step 2, Bob sends back the evaluation results of the function to Alice.
          If Alice decided to continue, Alice re-generates the number of clock cycles and sends it to Bob.
          Then, Bob performs the evaluation and goes back to step 1.
\end{enumerate}

This protocol is included in step 5 to 7 of the protocol flow of two-party SCO,
since the ciphertext of the termination flag is included in the encrypted result.
In our PoC implementation of VSP, the termination flag is (homomorphically) generated by the Instruction Decode stage of the processor.%

Note that if the adversary model is not honest, Bob can try to 
send the Bootstrapping Key,
which includes the encrypted secret key, or arbitrary ciphertexts to Alice
for decryption, pretending that the TLWE ciphertext is encrypting the termination flag.
As a result, to extend the threat model of VSP into a malicious setting, 
we need to ensure the existence of a decryption oracle and the malleability of the underlying FHE schemes are overcame. 
We point out that adopting IND-CCA1 FHE~\cite{Yasuda2018,10.1007/978-3-642-28496-0_4} in combined with Verifiable Computation~\cite{10.1007/978-3-642-40084-1_6} can be a candidate solution for VSP in a malicious setting, and is one of our future works.

\section{Design and Implementation of VSP}
\label{sec:implementation}

\begin{table*}[t]
 \caption{The Phases of VSP and the Associated Subcommands of the Command-Line Interface kvsp}
 \label{tab:kvsp}
 \begin{tabular}{cccccccc}
    \toprule
    Phase& Key Generation& Registration& Compilation& Encryption& Evaluation& Decryption& Resumption\\
    \midrule 
    Subcommand& kvsp-gen & kvsp-genbkey & kvsp-cc & kvsp-enc & kvsp-run & kvsp-dec&kvsp-resume\\
    Modules & (a) & (a) & (b), (c)&(c)&(a), (d)&(a)&(a), (d)\\
 \bottomrule
\end{tabular}
\end{table*}

In this section, we explain how we designed and implemented VSP~\cite{KVSP}.

\subsection{Design Goals}

The following three design goals are prioritized during the design of VSP.

{\bf\noindent (i) C compatibility}

Since it is obviously difficult to actually adopt a secure framework if the framework
is inconvenient to use,
we decided to support high-level program representations so that users can use 
VSP with ease.
There are two reasons why we chose the C language as our high-level representation. 
First,  C is one of the most widely used programming languages. 
Second, the C language is designed to be fast, where extensive optimizations
have been devoted into the optimization of C-based programs, e.g., 
the LLVM framework~\cite{LLVM:CGO04}. Therefore, with C support, users of VSP can have easy
access to efficient programs.

{\bf\noindent (ii) ISA Optimization}

Due to the high computational demand, the number of logic gates that can be evaluated in parallel ($\mathfrak{g}$) over TFHE is limited by the number of parallel processing capacity of the physical machine.
In VSP, the evaluation time of the circuit is proportional to the total number of gate 
count ($\mathfrak{t}$), as $\mathfrak{g}$ of a processor generally exceeds the parallel processing capacity of an ordinary desktop computer.
Since the ISA plays a key role in determining $\mathfrak{t}$ of the processor, 
we decided to design our custom ISA in such a way that the circuit of the 
processor can be minimized, while retaining C compatibility. 

{\bf\noindent (iii) Maximizing Parallelism}

As mentioned, the amount of parallelism in VSP (and generally in lattice-based
cryptography) exceeds the parallel processing capability of conventional desktop
computers. However, we point out that cloud vendors may have much more computational
resources available than a home computer.
To fully leverage the computational resources available
in data centers, we designed the processor architecture in VSP 
to have a pipelined structure, where the processor circuit is divided into
different pipeline stages that can be evaluated simultaneously. 
We assert that the pipeline technique does affect
the execution time when the physical machine does not have much parallel
processing capability. However, the runtime of VSP can be significantly
reduced by the pipeline technique when there are enough physical processor cores.

\subsection{The Architecture of VSP}

\begin{figure}[t]
  \centering
  \includegraphics[width=\linewidth]{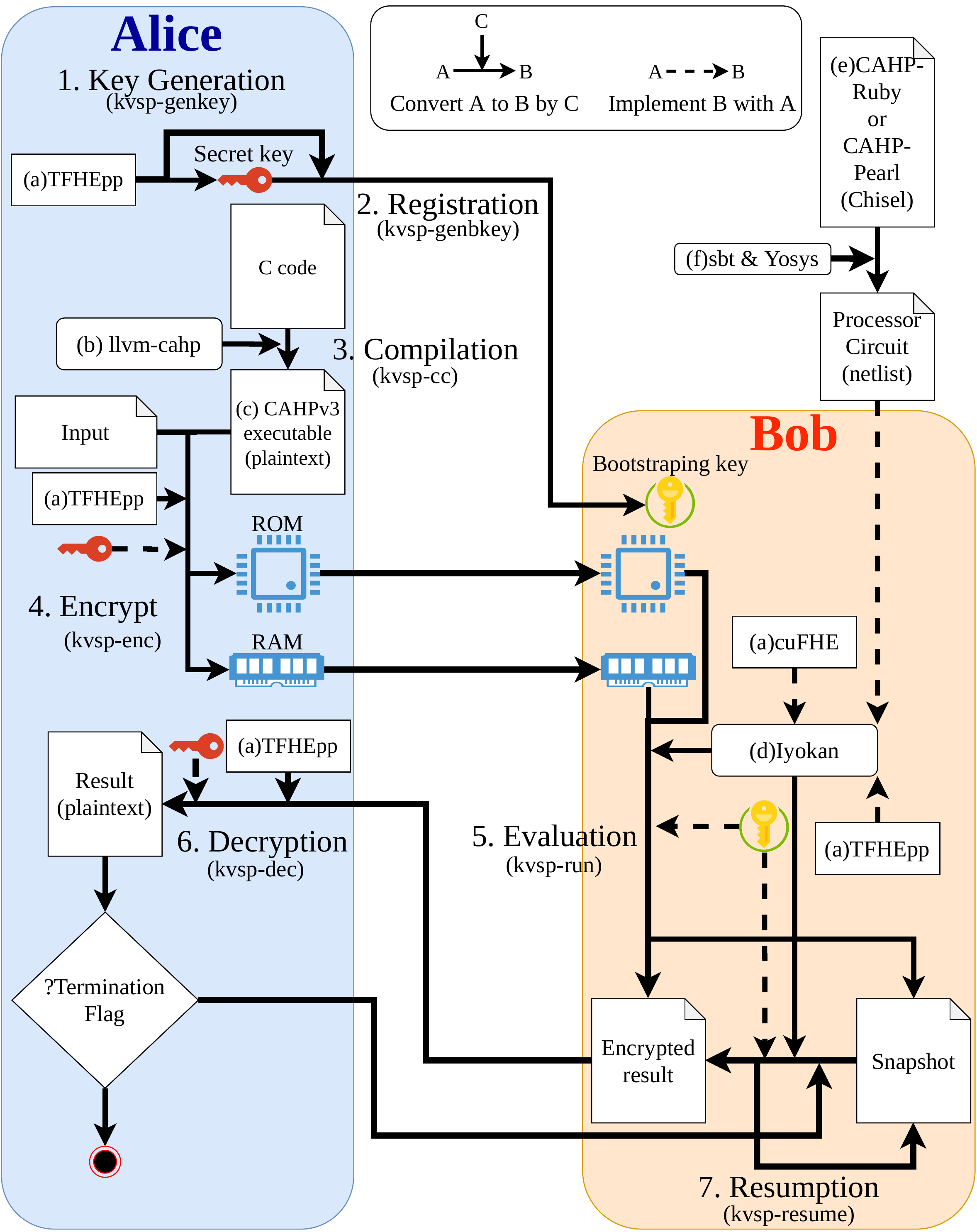}
  \caption{The VSP architecture and the main procedural flow.}
  \label{fig:vsparch}
\end{figure}

{\bf Notation}: In this paper, physical machine is the actual processing unit that runs VSP. CPU and GPU refers the physical CPU or GPU in the physical machine. In contrast, we use processor to refer to the virtual processor constructed over TFHE in VSP.

The visual overview of the implemented protocol flow of the proposed 
VSP framework is given in \cref{fig:vsparch}, which
details the abstract protocol flow in \cref{fig:scoworkflow}.
\cref{tab:kvsp} shows which phase each subcommand of kvsp~\cite{KVSP} (a command-line user interface for VSP) corresponds to and by which module is called.
Each subcommand of kvsp takes its inputs as a file, and outputs its results to a file. Therefore, the communication between the parties can be done via files transferring through public channels.

We first describe how the modules (a)-(f) are used here. Then, we explain each module.
In this work, we name our proposed processor circuits as (e)~CAHP-Ruby and CAHP-Pearl, and
the details on the circuits are explained in \cref{sec:proc}.
It is assumed that Alice and Bob agree on which processor architecture will be used in advance.
(f)~sbt~\cite{sbt} and Yosys~\cite{Yosys} are used to convert 
the Chisel code for the processor into a JSON netlist. Here, the netlist 
is a graph of nodes, where each node corresponds to a logic gate. The netlist is provided to Bob before the start of the protocol.
In Key Generation phase, Alice uses (a)~TFHEpp, a C++ implementation of TFHE on CPU, to generate a secret key.
In Registration phase, Alice uses (a)~TFHEpp one more times to generate the Bootstrapping Key from the secret key and sends the Bootstrapping Key to Bob.
In Compilation phase, Alice uses (b)~llvm-cahp, our C compiler for our custom ISA called (c)~CAHPv3, to generate executable binaries.
In Encryption phase, Alice uses (a)~TFHEpp to encrypt the executable binaries and the input into encrypted ROM and RAM. Then, Alice sends the ROM and RAM to Bob.
In Evaluation phase, Bob uses (d)~Iyokan to evaluate the processor circuit netlist over TFHE with the given encrypted ROM and RAM data. (a)~TFHEpp and cuFHE, the CUDA implementation of TFHE on GPU, are used in (d)~Iyokan to perform homomorphic computations. Then, Bob sends back the encrypted result to Alice.
In Decryption phase, (a)~TFHEpp is used to decrypt the encrypted result.
In Resumption phase, Alice checks the termination flag in the result. If it is 1, she terminates the protocol. Otherwise, Alice tells Bob to resume evaluation. Bob again runs (d)~Iyokan for the new round of homomorphic evaluation.

\subsubsection*{(a) TFHEpp and cuFHE: The TFHE libraries}

TFHEpp is our fully-scratch C++17 implementation of TFHE on the CPU, while
cuFHE is a TFHE library on the GPU (we optimized the original TFHE library 
from~\cite{cuFHE,10.1007/978-3-319-29172-7_11}).

In general, cuFHE is faster than TFHEpp, especially  when multiple logic 
gates are run in parallel, as the throughput of GPU is higher than that of CPU.
We describe how we use these libraries in (d)~Iyokan.

While TFHEpp supports Circuit Bootstrapping, which is a necessary component of CMUX Memory,
cuFHE does not. cuFHE uses Number Theoretic Transform (NTT) to perform fast polynomial 
multiplication,
where the ciphertext modulus
is kept to be $2^{64}-2^{32}+1$. This bit width constraint is to ensure
that the operands involved in NTT fit into the multiply instructions 
on the GPUs. Unfortunately, due to this bit width constraint, 
cuFHE cannot directly perform Circuit Bootstrapping, 
as the moduli required by Circuit Bootstrapping needs to be larger than 64-bit.
While we can simply increase the size of modulus to be compatible
with Circuit Boostrapping, the performance of cuFHE in practice will
be significantly reduced as more multiplication instructions (on the GPUs) and memory accesses are required to perform
a single polynomial multiplication operation. The efficient implementation
of Circuit Bootstrapping on GPUs currently remain as an open field of study.

\subsubsection*{(b) llvm-cahp: The C Compiler}

We implemented a new C compiler \textit{llvm-cahp} for our ISA, CAHPv3, using LLVM9.

The LLVM compiler infrastructure project is an assemblage of compiler and toolchain technologies~\cite{LLVM:CGO04},
which serves as a good foundation for our custom processor architecture and ISA.
LLVM is widely used in both open and closed projects as well as used in academia~\cite{LLVMUsers}. In particular, LLVM
surpasses GCC to win the ACM Software System Award in 2012~\cite{ACMLLVM}.
LLVM includes four parts. First, we have language-dependent frontends that compile the program source code into
the intermediate representation named LLVM IR. Second, LLVM has a target-independent optimizer that operates on LLVM IR.  Third, 
the LLVM target-specific backends are used to generate the object code of each 
target from LLVM IR. Finally, 
the LLVM linker turns multiple object codes into one executable.
Since we defined a custom ISA, we implemented a new backend for CAHPv3.
We also added support for CAHPv3 to the frontend of the C language (i.e., Clang), and to 
the LLVM linker (i.e., LLD).
By putting them together, we can directly compile C program into a CAHPv3 executable binary file.

Our compiler supports almost all features of C such as basic arithmetic operations, 
control expressions,
function calls including recursion, structures, and so on.
Furthermore, since LLVM has the target-independent optimizer as mentioned above,
llvm-cahp can output fast and small executables by using the \texttt{-O3} or \texttt{-Oz} compiler options.

Since the proposed processor is a virtual one, our modified compiler
does not provide functions in standard libraries that require
physical processor components (e.g., the print function). There
are also some minor limitations (e.g., jump over 1kiB) in our
compiler.

\subsubsection*{(c) CAHPv3: Instruction Set Architecture}

CAHPv3\footnote{CAHP is short for ``CAHP Ain't for Hardware Processors,''
and v3 means this is our third version ISA for VSP (the former two did not work well).}
is our RISC ISA based on RISC-V 32-bit integer and 16-bit compressed instructions (RV32IC).
CAHPv3 has 16-bit datapath and sixteen 16-bit registers. However, the instruction bit width is a mixture of 24 bits and 16 bits,
since we want to minimize the size of the machine code. 

CAHPv3 has two important features from the perspectives of our design goals.
First, it is relatively easy to implement the LLVM backend for CAHPv3, due
to its similarities to
mainstream ISAs such as x86 and RISC-V. We note that this is one of
the main reasons why the OISC used in FURISC is considered impractical.
Second, CAHPv3 reduces the complexity of the processor circuitry 
because it is a RISC ISA, and the datapath is only of 16-bit wide.
Unlike RV32IC, CAHPv3 does not include instructions that are not necessary in VSP,
such as privileged instructions and synchronization instructions, further
reducing the total gate count. The specification is here~\cite{specification}.

\subsubsection*{(d) Iyokan: The Gate Evaluation Engine}

Iyokan is our main software written in C++17 to run the processor over TFHE.
The fundamental features of Iyokan are to receive an arbitrary Boolean circuit along
with the encrypted input data,
evaluate the circuits according to the inputs over TFHE, and return 
encrypted results of the evaluation.
Therefore, we can execute encrypted programs without decryption by feeding Iyokan
with the processor as a logical circuit and the associated inputs.

Iyokan works in the following way:
\begin{enumerate}
    \item Split the input sequential logical circuit into two parts: combinational circuits
        and flip-flops to represent general Boolean circuits. %
    \item Convert the combinational circuits into a directed acyclic graph (DAG),
        where the logical gates are represented as graph nodes, and
        wires as directed edges.
        \cref{fig:graph} shows an example graph representation of the half adder circuit 
        in \cref{fig:halfadder}.
    \item Evaluate the DAG by using the converted circuit along with its inputs and
        the outputs of the flip-flops. 
        Since every node in the DAG has to be evaluated, Iyokan uses the 
        list scheduling algorithm
        to assign the tasks to workers which are physical CPU and GPU processing units.
        Note here that the scheduling algorithm also needs to resolve
        the dependency relations between nodes represented as edges in the DAG.
        Almost all the tasks are executed on GPUs via cuFHE, and
        the rest of the tasks which cannot be run on GPUs, such as Circuit Bootstrapping, are executed on CPUs via TFHEpp.

        This step gives us the output of the combinational circuit in the current cycle,
        which is used as the inputs to the flip-flops.
    \item Save the inputs the previous step provides to the flip-flops (physical memories).
    \item Output the stored values in the flip-flops.
    \item Exit if the number of clock cycles exceeds the threshold which is specified by the user through command-line option. 
    Otherwise, go to step 3.
\end{enumerate}

Each evaluation from step 3 to 6 corresponds to one clock cycle.
As mentioned in \cref{subsec:abstract-protocol-flow},
Alice has to decide a threshold, that is, how many times the steps between step 3 and 6 
should be repeated.

There are two important features of Iyokan. First, Iyokan can handle not only normal logic
gates but also CMUX Memory. CMUX Memory can be represented as a scheduled graph,
so it can also be embedded in the DAG. Second,
Iyokan can run more than one worker on CPUs and GPUs in parallel.

\subsubsection*{(e) CAHP-Ruby, CAHP-Pearl: Processor}
We developed two processors, CAHP-Ruby and CAHP-Pearl, for VSP:
\begin{description}
    \item[CAHP-Ruby] CAHP-Ruby is a 5-stage pipeline processor that implements CAHPv3 ISA.
    We will explain its details in \cref{sec:proc}.
    \item[CAHP-Pearl] CAHP-Pearl is a single cycle processor that also implements CAHPv3 ISA.
    We made it by just removing pipeline registers from CHAP-Ruby.
\end{description}

\subsubsection*{(f) sbt and Yosys: Logic Synthesis}

We chose Chisel~\cite{6241660}, a particular Hardware Description Language (HDL), to 
instantiate our processors for VSP, as Chisel is widely adopted in the industry~\cite{Chiselusage}. %
The sbt program compiles Chisel to the Verilog HDL.
Then, Yosys~\cite{Yosys} is utilized to compile Verilog codes into JSON netlists.

\section{The Proposed Processor Architecture}
\label{sec:proc}

\begin{figure}[t]
  \centering
  \includegraphics[width=\linewidth]{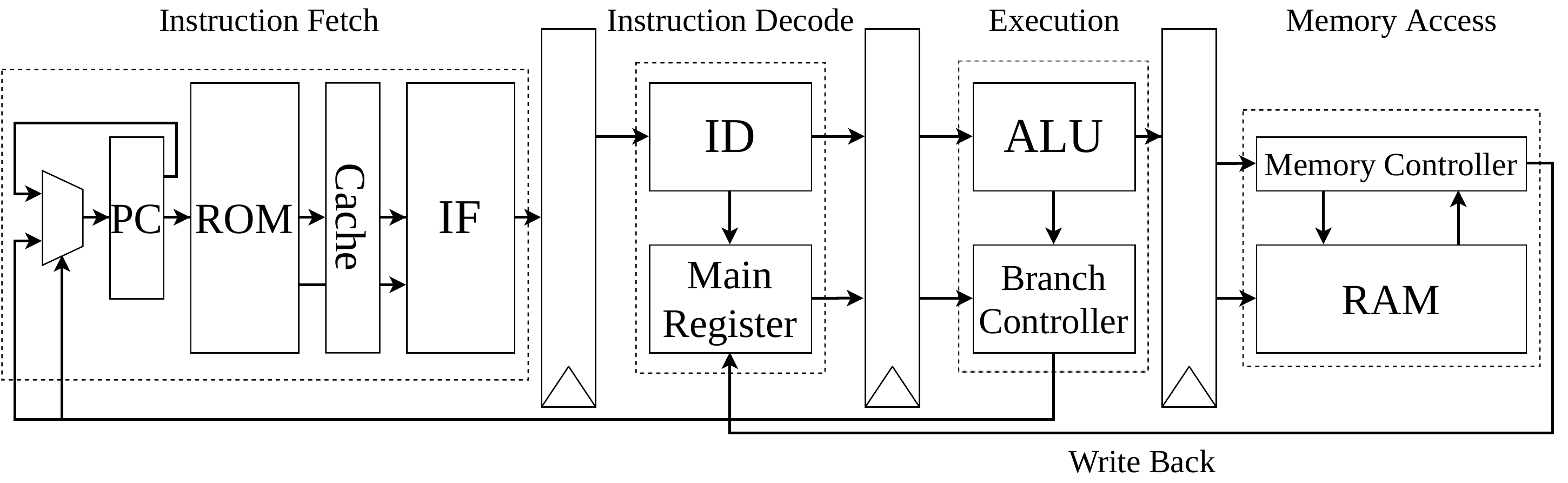}
  \caption{The architecture of the five-stage pipelined CAHP-Ruby processor.}
  \label{fig:ruby-arch}
\end{figure}

\cref{fig:ruby-arch} conceptually illustrates CAHP-Ruby, the proposed 
custom processor architecture. 
CAHP-Ruby has a five-stage pipeline structure consisting of an \textit{Instruction Fetch}, an \textit{Instruction Decode}, an \textit{Execution}, a {\it Memory Access}, and a \textit{Write Back} stage.
We chose a five-stage construction, as  
this structure is widely used in physical processor designs~\cite{10.5555/1502247,RocketChip,10.5555/546884,MicroBlaze}.
Determining the optimal number of pipeline is actually platform-dependent, i.e., 
it depends on the physical resources available to VSP. A framework that automatically optimizes 
the number of pipeline stage is one of our main future works.

CAHP-Ruby has two different memory areas: ROM and RAM, as shown in \cref{fig:ruby-arch}.
This structure greatly simplifies the processor circuitry and 
enables each memory area to have different and optimized implementations,
further discussed in \cref{sec:cmux}.
Here, ROM is a read-only memory area, designated for the compiled instructions. 
RAM permits both read and write operations, and is mainly for program data handling. 
We note that CAHP-Ruby does not support any peripheral devices nor interruption because 
they are not needed in a virtual processor.
Through such design decisions, we are able to reduce the complexity of the CAHP-Ruby circuitry.

In what follows, we detail the operational behavior of each of our custom
processor stages.

{\bf Instruction Fetch (IF)}: %
IF is responsible for producing an instruction.
First, IF fetches a 32-bit block from ROM. However, 
the block may not contain any complete instructions due to the fact that
our custom ISA contains both 16-bit and 24-bit instructions.
Therefore, IF includes a 32-bit instruction cache to resolve this 24/16-bit boundary alignment problem.
The cache contains ROM output value of the previous clock cycle. 
If the currently fetched 32-bit ROM block does not contain a complete instruction,
data from the instruction cache can be read, and it is guaranteed that there will
always be a complete instruction in a 64-bit ROM block.
Therefore, IF constructs a complete instruction with the assistance of the instruction
cache and the current ROM output value.

{\bf Instruction Decode (ID)}:
ID decodes the instruction to provide operands for the execution stage. %
This stage also reads the data from the registers specified by the instruction in
the main register file.
ID is also responsible for generating the termination flag. 
In this work, we indicate a program termination by inserting a jump instruction
which jumps to the same its own memory address, creating an infinite loop.
Once the ID stage detects such loop,  the termination flag is set, and
can be read from the dedicated port.

{\bf Execution (Ex)}:
This stage consists of an arithmetic and logical unit (ALU) and a branch controller.
ALU performs (homomorphic) arithmetic operations such as addition and subtraction, and
logical operations such as logical summation, and shift.
In the case of a jump instruction or a branch instruction, the branch controller 
generates a flag indicating whether to jump or not according to the result of the ALU 
operation. We assert that all the computations and branches are over FHE ciphertexts,
guaranteeing that the processor circuit evaluator does not observe 
any private information.

{\bf Memory Access (Mem)}:
This stage consists of two parts: memory controller and RAM.
We defer a detailed presentation of the RAM in 
Section~\ref{sec:cmux}.
The memory controller takes write data from the execution stage as its input.
When the write data is 8-bit wide, the controller converts the write data to be of 16-bit wide,
for the RAM only accepts 16-bit data.
The memory controller also reads the data from the read port of RAM and
format when the output value to be of 8-bit wide. Finally, the memory
controller passes the read data from the RAM to the write back stage.

{\bf Write Back (WB)}:
This stage simply writes data into the main register files.

\section{CMUX Memory}\label{sec:cmux}

In this section, we present CMUX Memory, a new construction of memory unit over HE that leverages the LHE mode of TFHE for optimization. 
As mentioned, there are two types of memories: RAM and ROM.

\subsection{Theoretical Speed Predictions}
Informally, the reason why CMUX Memory is fast can be explained by the fact that the 
evaluation of Circuit Bootstrapping takes about 10 times as long as it takes to evaluate 
any two-input homomorphic gate. %
Let $v,w \in \mathbb{N}$ be the number of bits of the address and the data bus, respectively.
Assuming that we ignore the time it takes to process CMUX, because CMUX is several hundred times faster than any two-input homomorphic gate, the time it requires to evaluate the ROM of the CMUX Memory is roughly equivalent to $10v+w$ 
Homomorphic Gates. Meanwhile, the time it takes to evaluate the RAM is roughly equivalent to 
$10v + w(2^v+1)$ Homomorphic Gates. The $w$ term comes from HomMUX without SE and IKS and the $w\cdot2^v$ term comes from the noise 
refreshment.
The construction of the ROM and RAM by logic gates takes at least $w\cdot2(2^v-1)$ two-input Homomorphic Gates each to construct the tree for data fetching. 
Therefore, in theory, CMUX Memory can be expected to be faster than constructing the memory by logic gates.

\subsection{RAM}

\begin{figure}[t]
  \centering
  \includegraphics[width=0.9\linewidth]{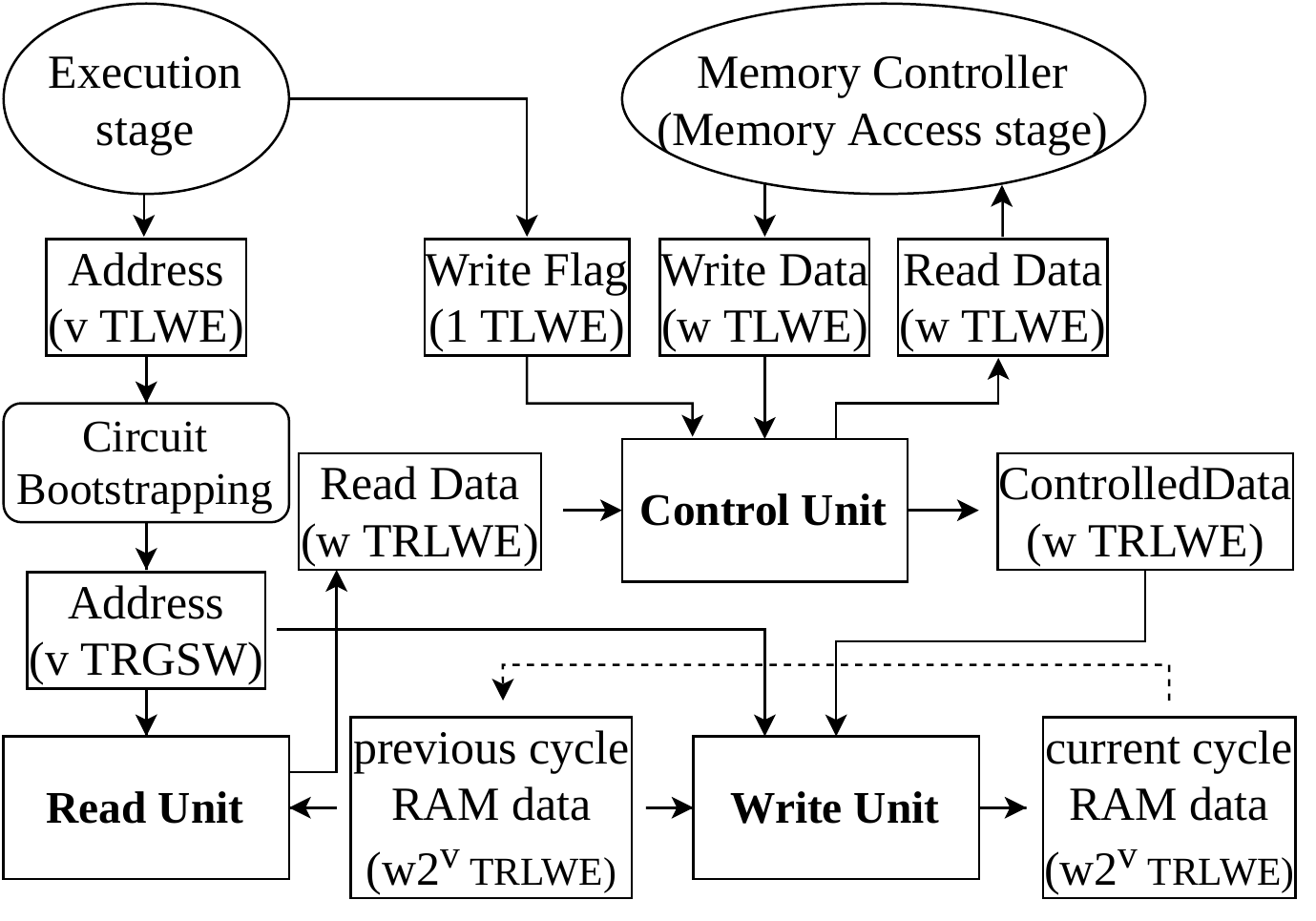}
  \caption{The architecture of the one-cycle single-port RAM.}
  \label{fig:ramarch}
\end{figure}

In this paper, we only treat one cycle, single port RAM since it requires
the minimum amount of Bootstrapping to implement. %
The RAM has following characteristics: (i) Read and write are exclusive. (ii) Both read and write are done in one cycle. (iii) Read and write use the same address.

The visual overview of the RAM architecture is given in \cref{fig:ramarch}.
There are three inputs for RAM: address, write flag, and write data. 
The address is the memory address for write or read.
The write flag is one bit data which selects the operation mode of RAM. 
The write data is the data which will be wrote in the address if the RAM is write mode.
The RAM has one output port, where the data presented at the input address are retrieved.
In VSP, the data bus in the processor uses TLWE as ciphertexts for memory elements, 
since TLWE ciphertexts are also used by the Homomorphic Gates in other parts of the
processor circuit.

As shown in \cref{fig:ramarch}, RAM consists of the read unit, the control unit, 
and the write unit. In what follows, we provide a comprehensive explanation
on each of the unit. Note that
addresses in the write and the read units are in TRGSW ciphertext, but the memory controller in Memory Access stage of the processor feeds the address as TLWE ciphertexts.
Therefore, Circuit Bootstrapping is applied to TLWE ciphertexts 
to get TRGSW ciphertexts representations of the addresses.

\subsubsection*{Read Unit}

\begin{figure}[t]
  \centering
  \begin{minipage}[t]{0.48\linewidth}
  \includegraphics[width=\hsize]{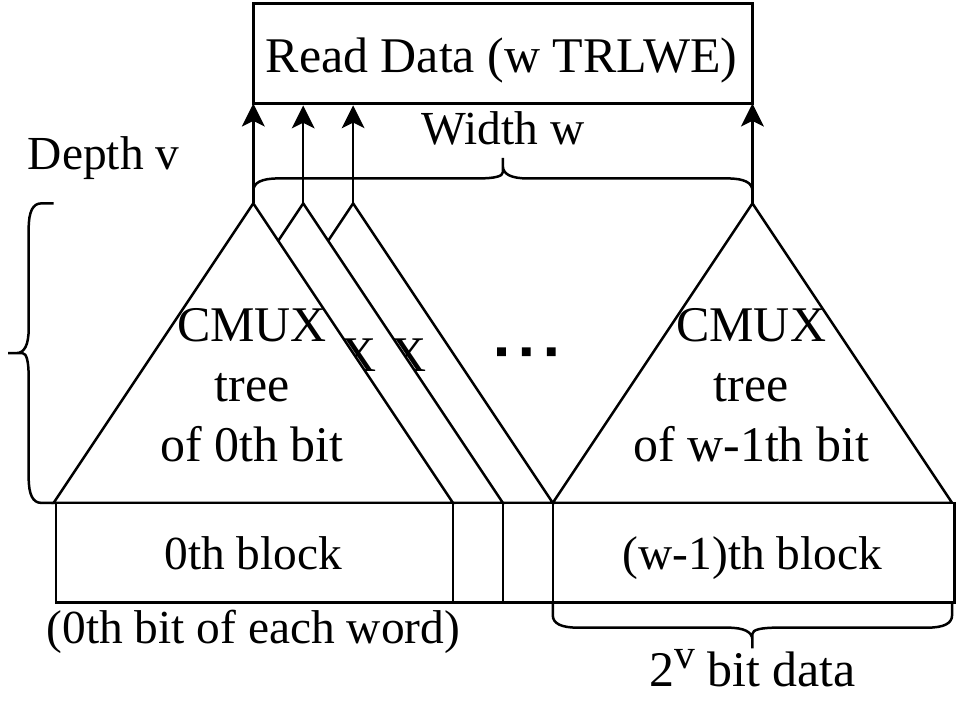}
  \caption{The architecture of the read unit.}
  \label{fig:readarch}
  \end{minipage}
  \begin{minipage}[t]{0.48\linewidth}
  \includegraphics[width=\hsize]{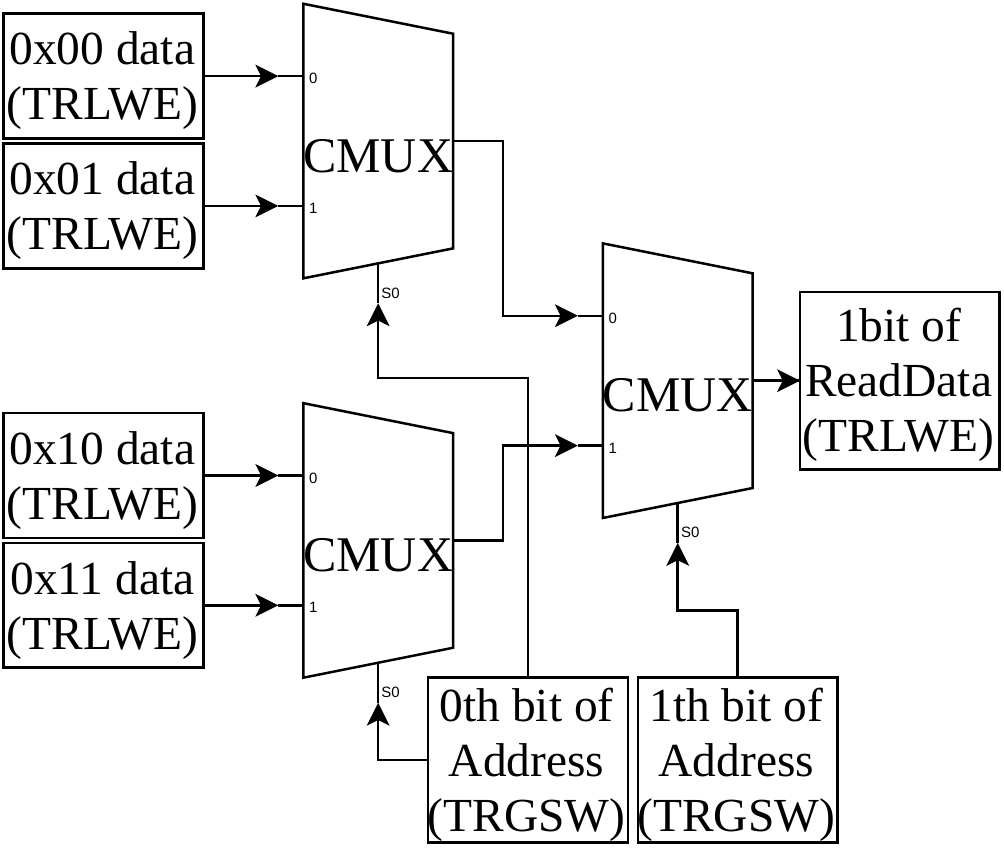}
  \caption{An example of the CMUX tree ($v=2$).}
  \label{fig:cmuxtree}
  \end{minipage}
\end{figure}

The read unit reads the data at a given address. The visual overview of its architecture is given in \cref{fig:readarch}~and~\ref{fig:cmuxtree}. 
The data of RAM are represented as $w\cdot 2^v$ TRLWE ciphertexts, where each TRLWE 
ciphertext contains one bit of plaintext information. 
The TRLWE ciphertexts are divided into $w$ blocks, where the $i$th block contains the $i$th bit of each word. A CMUX tree 
is used to fetch the $i$th bit of the word from the $i$th block.
We note that, although the message space of TRLWE is capable of holding a vector of
$N$ binary values, i.e., $\mathbb{B}_N[X]$, we only fill one entry with an actual
plaintext value.
If we pack multiple bits into a single TRLWE ciphertext for read, we also have to write in a packed manner. 
The problem for pack writing is that every instruction might have a chance to write only a small amount (e.g,. a 16-bit register) of data to RAM, and the amount of computations it takes to pack and unpack bits can be a significant overhead.

The task of the CMUX tree is to compare each bit of the address with that of 
RAM data by a tree of multiplexers implemented by CMUX, such that data at the
designated memory address can be read.

\subsubsection*{Control Unit}

\begin{figure}[t]
  \centering
  \includegraphics[width=0.7\linewidth]{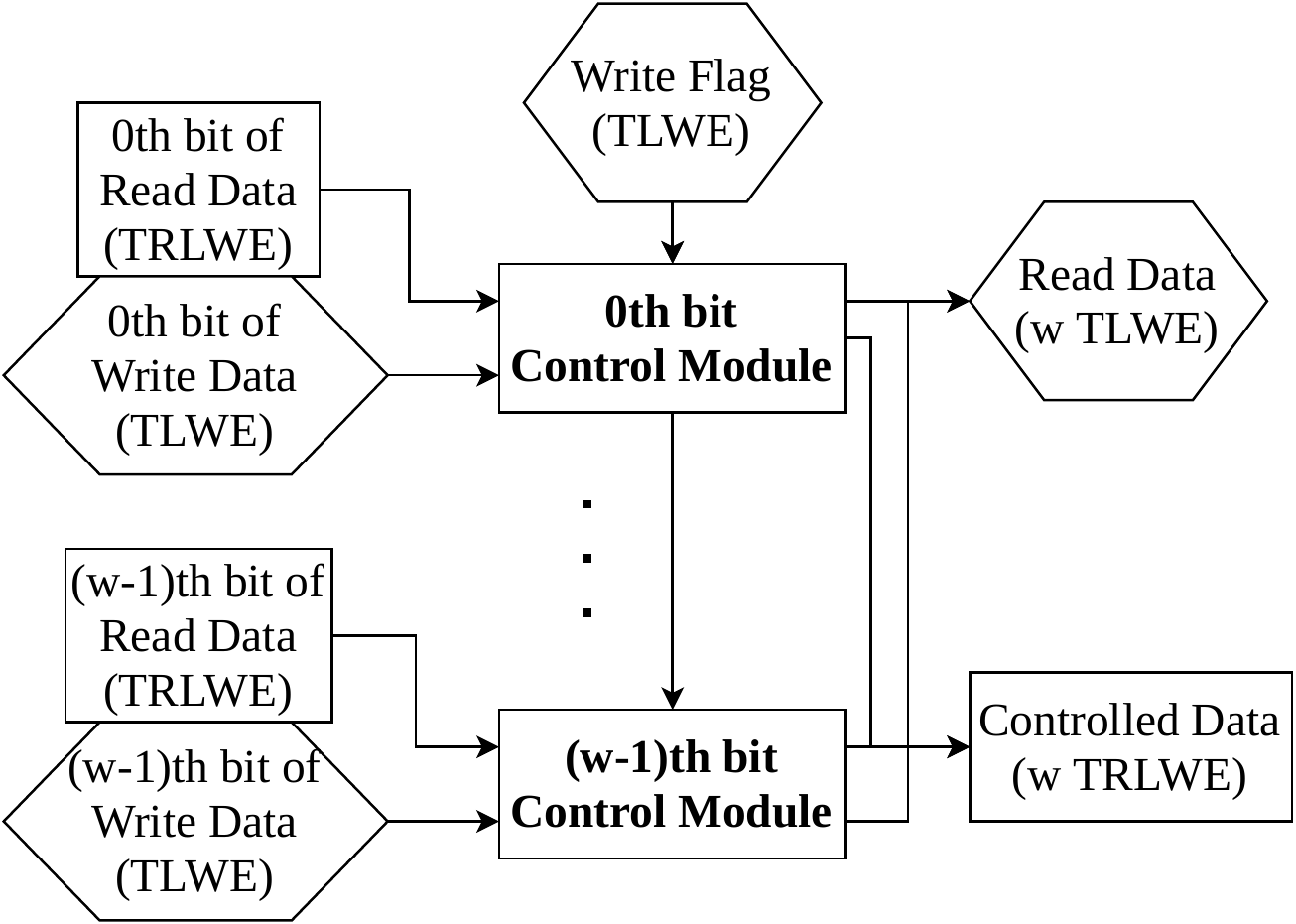}
  \caption{The architecture of the control unit.}
  \label{fig:ctrlarch}
\end{figure}

\begin{figure}[t]
  \centering
  \includegraphics[width=\linewidth]{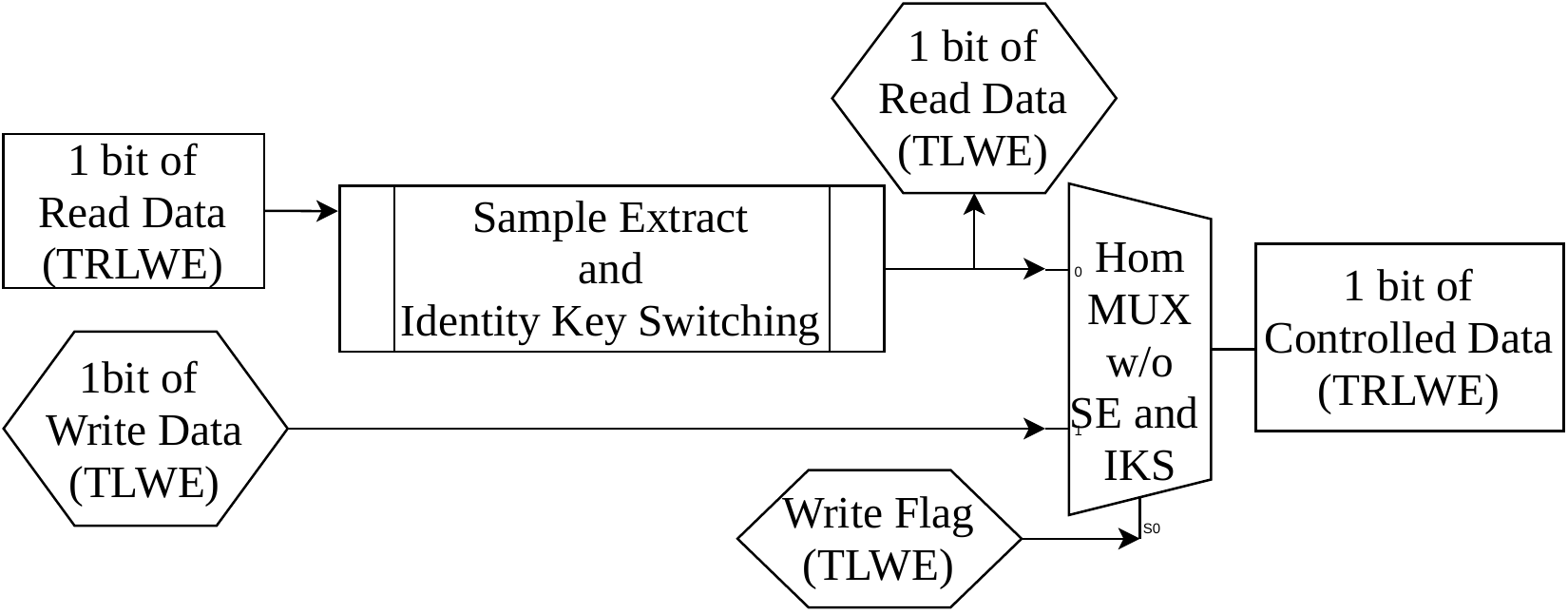}
  \caption{The architecture of a control module.}
  \label{fig:modulearch}
\end{figure}

The control unit is the interface between main processor circuit
and CMUX Memory.
We show an architectural illustration of the control unit and
module in  \cref{fig:ctrlarch}~and~\ref{fig:modulearch}, respectively.
The control unit consists of $w$ control modules, each of
which processes a single bit of the write data. 
Since the processor only accepts TLWE ciphertexts, SE and IKS 
are inserted to convert the read data from the read unit into TLWE ciphertexts. The control module performs multiplexing between the read and the write 
data, depending on the write flag. 
The multiplexed result is sent to the write unit as the controlled data.

\subsubsection*{Write Unit}

\begin{figure}[t]
    \centering
    \includegraphics[width=0.6\linewidth]{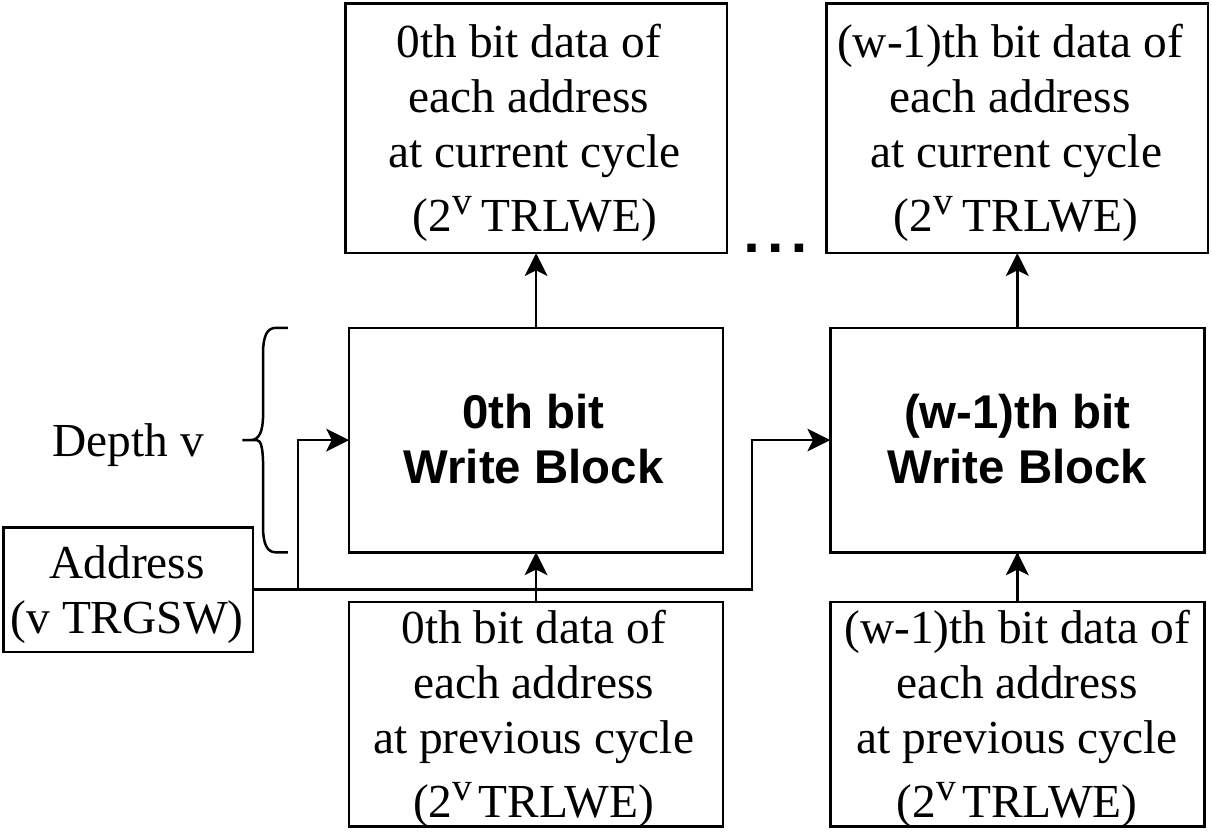}
    \caption{The architecture of the write unit.}
    \label{fig:writearch}
\end{figure}

\begin{figure}[t]
    \centering
    \includegraphics[width=0.6\linewidth]{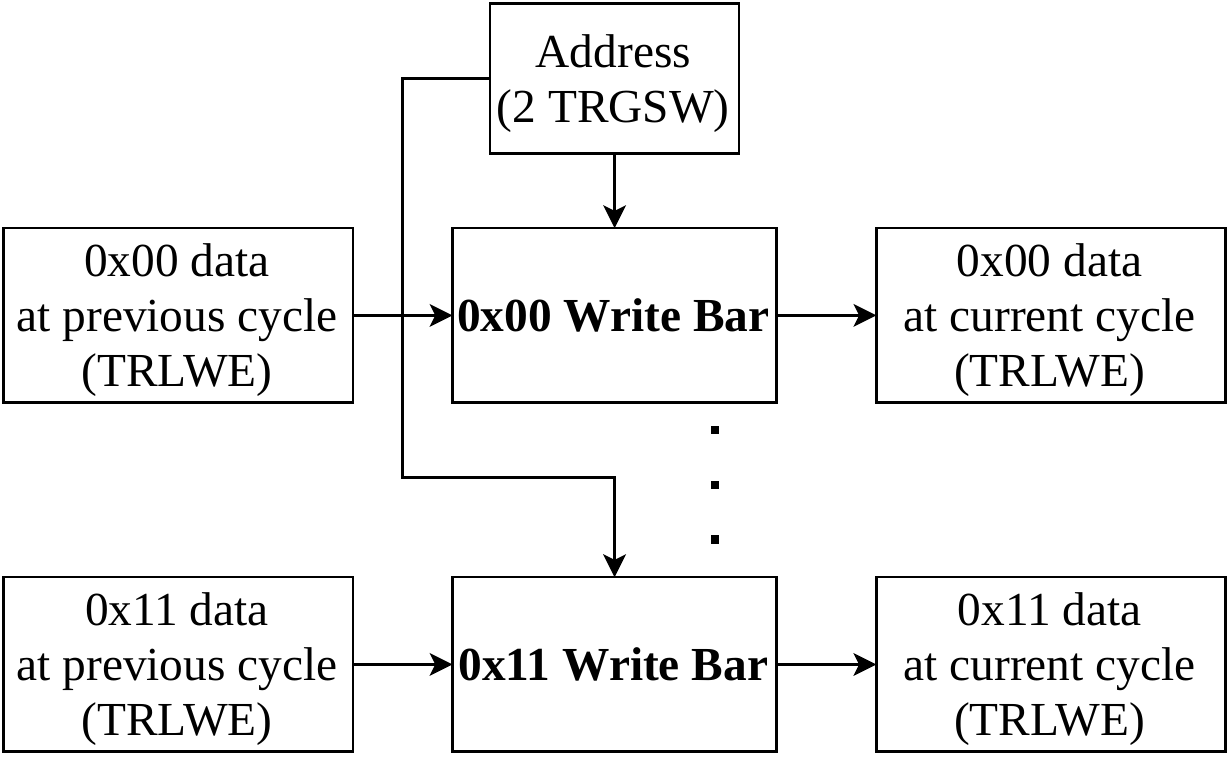}
    \caption{An example of the write block ($v=2$).}
    \label{fig:blockarch}
\end{figure}

\begin{figure}[t]
  \centering
  \includegraphics[width=0.6\linewidth]{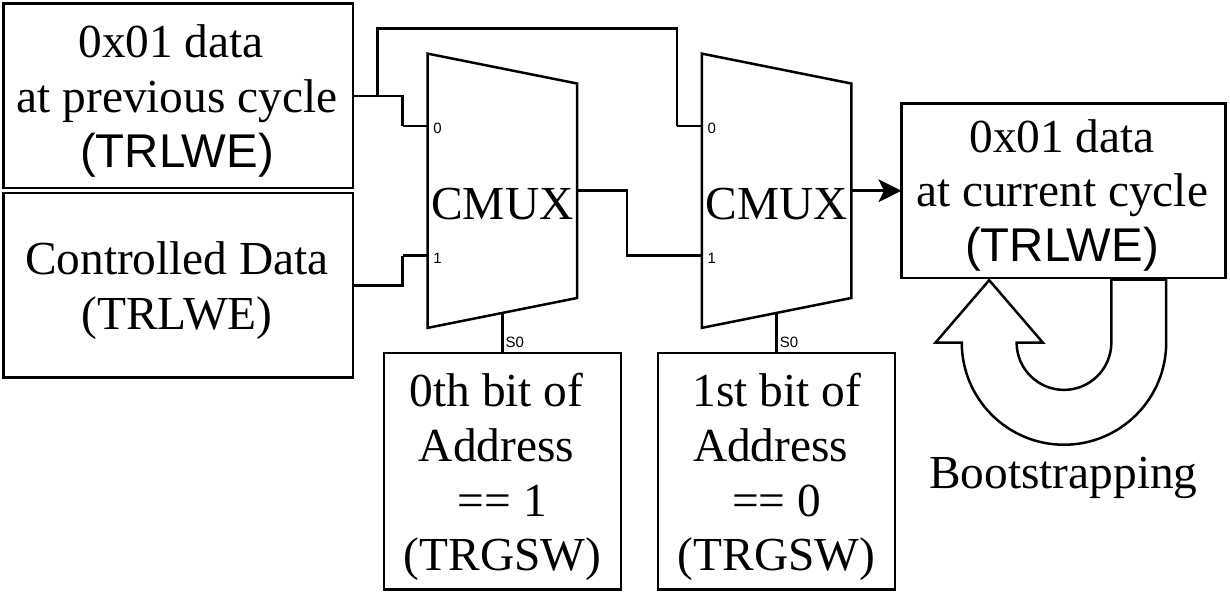}
  \caption{An example of a write bar at address 0x01 ($v=2$).}
  \label{fig:bararch}
\end{figure}

From the view of the main processor circuit, each word of the current cycle data is the multiplexed result between the word of the previous cycle data and the write data depending on the write flag and the address matching.
Since the multiplexed result depends on the write flag that is fed as the controlled data, the write unit only needs to take care of the address matching part of the computation.
The write unit also performs Bootstrapping over the entire contents of the RAM.
An visual overview of the write unit is given in \cref{fig:writearch},~\ref{fig:blockarch},~and~\ref{fig:bararch}.
The write unit consists of $w$ write blocks, each handles a single word.
Each write block 
is composed of $2^{v}$ write bars which handles a single bit.
Therefore, the write unit consists of $w\cdot2^{v}$ write bars arranged in parallel.

The working principle of the write bar is comparing each bit of the input address with the addresses in the RAM, through an array of CMUX gates.
If all bits in the input address match a particular entry in the RAM, the controlled data is selected and becomes current cycle data. 
Here, when the write flag is false, the controlled data is same as the previous cycle data in the address, so current cycle data is same as previous one.
On the other hand, if the write flag is true, the controlled data and current cycle data
both become the write data, and the data are written into the memory.
If the addresses do not match, previous cycle data is selected, and data in memory are not modified. The write bar refreshes the noise added by the CMUX array by bootstrapping the data at the end.

{\bf Remark}: The implementation of the comparison between an input address 
bit and a constant address bit is, in fact, quite simple. 
More specifically, the comparison result between an input
bit with a constant value of 1 is the bit itself. Meanwhile, 
the comparison with 0 
can be implemented by a 
subtraction of a constant TRGSW ciphertext encrypting 
the constant 1 followed by a sign inversion of all coefficients in the resulting TRGSW ciphertext.

\subsection{ROM}

The construction of ROM with LHE mode of TFHE is trivial by using Look Up Table (LUT), which is described in~\cite{Chillotti2020}. We applied both optimization techniques mentioned in~\cite{Chillotti2020}, namely Vertical Packing and Horizontal Packing.

\section{Evaluation}
\label{sec:eval}

In this section, we perform thorough experiments on VSP to demonstrate
its performance. We will first characterize VSP over a set of 
benchmarks in Section~\ref{sec:exp_bench}, and then deliver the overall
performance statistics in Section~\ref{sec:exp_overall}

\subsection{Benchmarks}\label{sec:exp_bench}

\begin{table}[t]
  \centering
  \caption{Processor Size Evaluation}
  \label{tab:circuitsize}
  \small
  \begin{tabular}{c|ccc}
        \toprule
       Processor & MUX & NOT & Others  \\
       \midrule
       CAHP-Ruby & 996 & 15 & 2422\\
       CAHP-Pearl & 877 & 22 & 2054\\
       Lite MIPS~\cite{Lite_MIPS}& 1276& 39& 6241\\
       PicoRV32~\cite{PicoRV32} & 2732& 11& 5162\\

       \bottomrule
 \end{tabular}
\vspace{-3mm}
\end{table}

\begin{table}[t]
  \centering
  \caption{Size of Keys and Ciphertexts}
  \label{tab:size}
  \begin{tabular}{cc}
    \toprule
    Type&Size[MiB]\\
    \midrule
    Secret Key & 0.023\\
    Bootstrapping Key & 2563.047 \\ %
    ROM & 0.033 \\
    RAM & 33.55\\
  \bottomrule
\end{tabular}
\end{table}

\begin{table}[t]
  \centering
  \caption{Machine Code Size}
  \label{tab:codesize}
  \begin{tabular}{ccc}
    \toprule
    Program&RV32IC [B]&CAHPv3 [B]\\
    \midrule 
    Fibonacci & 36 & 31 \\
    Hamming & 354 & 264 \\
    Brainf*ck & 226 & 229 \\
  \bottomrule
\end{tabular}
\end{table}

\begin{table*}[t]
  \caption{Performance Evaluation Using Hamming} %
  \label{tab:program-evaluation}
  \small\centering
  \begin{tabular}{c|cccc|cccc}
      \toprule
      Case \# & Machine & \begin{tabular}{@{}c@{}}\# of\\V100\end{tabular} & Pipelining? & CMUX Memory? & \begin{tabular}{@{}c@{}}\# of\\cycles\end{tabular} & Runtime~[s] & sec./cycle & \begin{tabular}{@{}c@{}}\# of\\tries\end{tabular}\\
      \midrule
      1 & \multirow{2}{*}{AWS c5.metal} & \multirow{2}{*}{0} & No & Yes & 936 & $2342.0 \pm 13.3$ & $2.502 \pm 0.014$ & 3 \\
      2 &                               &                    & Yes & Yes & 1216 & $2773.0 \pm 2.8$ & $2.280 \pm 0.002$ & 3 \\\hline%

      3 & \multirow{4}{*}{Sakura Koukaryoku} & \multirow{4}{*}{1} & No & No & 936 & $5919.0 \pm 33.1$ & $6.324 \pm 0.035$ & 5 \\
      4 &                                    &                    & No & Yes & 936 & $2232.1 \pm 1.7$ & $2.385 \pm 0.002$ & 5 \\
      5 &                                    &                    & Yes & No & 1216 & $7809.0 \pm 45.8$ & $6.422 \pm 0.038$ & 4 \\
      6 &                                    &                    & Yes & Yes & 1216 & $2045.0 \pm 4.6$ & $1.682 \pm 0.004$ & 5 \\\hline%

      7 & \multirow{2}{*}{AWS p3.8xlarge} & \multirow{2}{*}{4} & No & Yes & 936 & $1455.7 \pm 0.3$ & $1.555 \pm 0.000$ & 3 \\
      8 &                                 &                    & Yes & Yes & 1216 & $979.0 \pm 12.5$ & $0.805 \pm 0.010$ & 3 \\\hline%

      9 & \multirow{4}{*}{AWS p3.16xlarge} & \multirow{4}{*}{8} & No & No & 936 & $1627.0 \pm 4.2$ & $1.739 \pm 0.004$ & 3 \\
     10 &                                  &                    & No & Yes & 936 & $1440.0 \pm 2.5$ & $1.538 \pm 0.003$ & 3 \\
     11 &                                  &                    & Yes & No & 1216 & $1566.0 \pm 9.7$ & $1.288 \pm 0.008$ & 3 \\
     12 &                                  &                    & Yes & Yes & 1216 & $\mathbf{965.9 \pm 3.4}$ & $\mathbf{0.794 \pm 0.003}$ & 3\\
      \bottomrule
\end{tabular}
\vspace{-3mm}
\end{table*}

\subsubsection*{Benchmark environments}

In our implementation, ROM and RAM are 512 bytes,
that is, $v=8$ and $w=16$ when using the CMUX Memory for RAM.
We also experimented 1 KiB ones. See \cref{sec:fulleval} for the details.

The main benchmark program used in our evaluation is Hamming.
Hamming takes two 8-digit hexadecimal numbers $a$ and $b$ as its arguments,
and finds the Hamming distance between them.
We include more benchmark programs (Fibonacci and Brainf*ck) in \cref{sec:fulleval},
and only point out that the general trends are the same as Hamming.
The programs are compiled into CAHPv3 executable by llvm-cahp with \verb|-Oz| optimization flag,
which minimizes the size of machine code. Then, the compiled programs
are encrypted and executed on Iyokan with CAHP-Ruby (with pipeline) and CAHP-Pearl (without pipeline).
The scripts to reproduce the runtime performance evaluation is available at~\cite{benchmark}.

We used four types of machines to evaluate VSP:
\begin{description}
    \item[AWS c5.metal]
        An HPC server hosted by Amazon Web Service equipped with
        Intel Xeon Platinum 8275CL CPU (96 vCPUs), 92GiB RAM, and no GPUs.
    \item[Sakura Koukaryoku]
        An HPC server hosted by Sakura internet Inc. equipped with
        Intel Xeon CPU E5-2623 v3 (16 vCPUs), 128GB RAM, and single NVIDIA Tesla V100.
    \item[AWS p3.8xlarge]
        An HPC server hosted by Amazon Web Service equipped with
        Intel Xeon CPU E5-2686 v4 (32 vCPUs), 244GB RAM, and 4 NVIDIA Tesla V100.
    \item[AWS p3.16xlarge]
        An HPC server hosted by Amazon Web Service equipped with
        Intel Xeon CPU E5-2686 v4 (64 vCPUs), 488GB RAM, and 8 NVIDIA Tesla V100.
\end{description}

\subsubsection*{Runtime Performance Evaluation}

\cref{tab:program-evaluation} shows the run-time statistics
required to evaluate the encrypted program of Hamming. %
Here, sec./cycle stands for seconds per clock cycle,
which is the amount of program run-time divided by the number of required clock cycles.%

While pipelining increases the number of gates of the processors,
the technique enables more gates to be run in parallel. 
Therefore, when the physical machine
has enough parallel processing units, pipelining reduces per-clock-cycle run-time 
of VSP, and eventually results in decreased total run-time (Compared between Cases \#4 and 6, 7 and 8, 9 and 11, and 10 and 12).
On the other hand, when the physical machine is not so powerful (Cases \#1 and 2),
the runtime ends up being slower due to the increased number of clock cycles.
In addition, in Cases \#3 and 5, the CMUX Memory is turned off and
the machine does not have enough parallel processing units to fully 
parallelize the gates in ROM and RAM. Consequently, the physical 
processors do not have more
machine resources for evaluating the pipelined 
core processor circuit.

Finally, we observe that while AWS p3.8xlarge (4 V100) is much faster than Sakura Koukaryoku (single V100),
there is almost no difference between AWS p3.16xlarge (8 V100) and p3.8xlarge.
This is most likely caused by the fact that the parallel processing capabilities of both machines
well exceed the number of logic gates that can be evaluated in parallel in our processor. Therefore, further pipelining may be conducted on such powerful
computing platforms.

Besides pipelining, we also experiment on the performance impact of
the proposed CMUX Memory. As shown in Table~\ref{tab:program-evaluation}, 
CMUX Memory reduces runtime across all cases we tested.
When CMUX Memory is not used, ROM and RAM need to be implemented 
by the Homomorphic Gates in the FHE mode of TFHE,
which results in significant performance degradations.

The fastest instance we tested is Case \#12, that is, AWS p3.16xlarge with pipelining and CMUX Memory applied,
which is shown in bold in \cref{tab:program-evaluation}. We achieved a performance of about 0.8~sec./cycle, or equivalently, 1.25~Hz.
From the results of the benchmark,
we conclude that both pipelining and CMUX Memory are effective in improving the performance of VSP.

\subsubsection*{Processor Size Evaluation}
In general, fewer logic gates means fewer computational complexity,
so the total gate count of the processors is one of the most important factors
which determine the performance of VSP.
\cref{tab:circuitsize} shows the size of CAHP-Ruby and CAHP-Pearl.
In the table MUX and NOT are counted separately
because their performance characteristics are different from a normal homomorphic gate.
In particular, MUX is twice as slow as other homomorphic gates, even
with the cryptographic optimization proposed in~\cite{Chillotti2020}.
Meanwhile, NOT can be evaluated much faster than other gates,
as the only operations in a NOT gate are sign inversions.
We compare the gate count of our processor to that of Lite MIPS~\cite{Lite_MIPS} and PicoRV32~\cite{PicoRV32}.
Lite MIPS is the processor which is implemented in TinyGarble~\cite{7163039}.
PicoRV32 is one of open-source implementations of RISC V, where the design goal
is to implement a small (in terms of gate count) processor.
As shown in \cref{tab:circuitsize}, our processors are smaller than both of the
existing designs.

\subsubsection*{Data size Evaluation}

We used two more programs except Hamming: Fibonacci and Brainf*ck here.
Fibonacci takes a decimal digit $n$ as its command-line argument, and calculates $n$th Fibonacci number.
Here we used $n=5$.
Brainf*ck interprets code of brainf*ck, which is a esoteric programming language, and returns the result.
We inputted \verb|++++[>++++++++++<-]>++| to it, the result of which is 42. 

\cref{tab:size} shows the size of keys and ciphertexts. We can see Bootstrapping Key is significantly bigger than other parts, so reusing Bootstrapping Key can reduce communication cost greatly. The reason why RAM is about 1024 times bigger than ROM is that Vertical Packing~\cite{Chillotti2020} is not applied to RAM. 
\cref{tab:codesize} shows the machine code size of the programs in CAHPv3.
We also show RV32IC version for reference.
RV32IC has more registers than CAHPv3 does, so register spills more often occurs in CAHPv3,
which made code of Brainf*ck larger.

\subsubsection*{Client-side Cost Evaluation}

It is noted that, on p3.8xlarge, it takes Alice (i.e., the client) 
about 57 seconds 
to complete the  generation of  the Bootstrapping Key, encryption of the 
memory, compilation of the program and the decryption of the 
evaluation results. Among these, the generation of 
Bootstrapping Key is the most time consuming procedure, where 
it takes about 55 seconds to finish. For simple programs like Hamming,
evaluating the program locally by the client only takes
around 0.5 microseconds on a conventional CPU, and program outsourcing in such 
case provides no practical merit.
However, as discussed in Section~\ref{subsec:abstract-protocol-flow}, 
for programs that potentially 
contain infinite loops, VSP can obviously reduce the amount of client-side
computations. Therefore, exploring practical applications of VSP is one of our main future works.

\subsection{Overall Performance and Comparison to Existing Works}\label{sec:exp_overall}
\begin{table}[t]
    \caption{Comparison between VSP and FURISC}
    \centering\small
    \begin{tabular}{c|ccc}
    \toprule
         Name & sec./cycle & \# of instructions & Implementation\\
         \midrule
         VSP &0.8 & Small&Public\cite{KVSP}\\
         FURISC & 1278 (est.)& Large&Private\\
         \bottomrule
    \end{tabular}
    \label{tab:compairison}
\vspace{-3mm}
\end{table}

Because FURISC~\cite{cryptoeprint:2015:699,Chatterjee2019} is the only previous work which represents the processor as a Boolean circuit and evaluates it over FHE,
we compare FURISC as the-state-of-the-art to VSP %
in \cref{tab:compairison}. FURISC gives the 
FPGA-accelerated evaluation time for 
Subtraction, in Table 6.5 in~\cite{Chatterjee2019}. 
Because SBN is the only instruction FURISC supports, 
the evaluation time of Subtraction corresponds to one clock cycle in the FURISC processor. 
Therefore, the estimated time for evaluating one clock cycle of FURISC is 
21.3 minutes, over 1000 seconds.
In contrast, our VSP implementation can evaluate one clock cycle in 0.8 seconds, as shown
in Case \#12 in \cref{tab:program-evaluation}.

The number of instructions for representing (almost all) programs in FURISC are larger than that of VSP,
for that FURISC has an OISC ISA. Therefore, we can see that compiling the same program
on VSP results in a smaller number of instructions, and that each instruction runs nearly $1600\times$ faster than FURISC. Hence, we are confident that the open-source VSP is the 
fastest FHE-based SCO platform to date.

\section{Conclusion}

In this work, we presented VSP,
the first comprehensive platform that implements a multi-opcode general-purpose sequential processor over TFHE for two-party Secure Computation Offloading (SCO).
We proposed a complete SCO scheme and designed a custom five-stage pipelined 
processor along with a custom ISA CAHPv3. 
We also proposed CMUX Memory, the optimized structure of ROM and RAM over TFHE to speed up instruction
evaluation.
We thoroughly evaluated VSP on benchmarks to show that both pipelining and CMUX Memory 
are effective in speeding up VSP.
Our open-source implementation is nearly $1600\times$ faster than the-state-of-the-art implementation while accepting conventional
C language programs. Therefore, we consider VSP as the first practical realization of a
two-party SCO scheme.

\section{Acknowledgement}

We thank all the people including our shepherd Thomas Ristenpart and anonymous reviewers
for their insightful comments.
This study was supported by Information-technology Promotion Agency (IPA), Japan,
The MITOU Program in fiscal year 2019, and SAKURA internet Inc.
We are grateful to Kazuyuki Shudo for his support as our project manager in The MITOU Program.
This work was partially supported by JSPS KAKENHI Grant No.~20K19799,
and 20H04156.
\bibliographystyle{ieeetr}
\bibliography{bibtex/vsp-acmccs.bib,bibtex/gc.bib,bibtex/he.bib,bibtex/homproc.bib,bibtex/oram.bib,bibtex/proc.bib,bibtex/sgx.bib,bibtex/uc.bib,bibtex/obfuscation.bib,bibtex/blind.bib}
\FloatBarrier
\appendix

\section{Related Work}\label{sec:additionalrealted}
In this section, we give some additional related works.

\subsection{Processor over HE}
Although FURISC is the most similar existing study to our proposed method, we will discuss other studies here.
There are some previous works which uses Paillier Cryptosystem~\cite{10.1007/3-540-48910-X_16} to evaluate encrypted binaries. HEROIC~\cite{6800460} is the first one. Paillier cryptosystem is one kind of Partial Homomorphic Encryption (PHE). Paillier Cryptosystem only permits addition of integers. Therefore, HEROIC uses some tables to provide enough functionality to implement a processor as Arithmetic circuit.
HEROIC implements an OISC processor which only supports SUBtract and branch if Less-than or EQual to zero (SUBLEQ) instruction. 
Unlike to FURSIC, there is a C like language compiler for SUBLEQ, HIGHER SUBLEQ, though its last update is in March 2011 \cite{highersubleq}. 
The use of the tables leads the ciphertexts becomes deterministic. This means the public key cannot be public.
Therefore HEROIC theoretically cannot achieve two-party PF-SFE.
In addition, the method of using the table has not been proven to be secure.
The authors of HEROIC also proposed Cryptoleq in 2016~\cite{7469876}. It also uses Paillier Cryptosystem with some tables and OISC.
They also proposed assembly-like Domain Specific Language (DSL) in it. Cryptoleq depends on the random number generation of the server. This is not suitable characteristic for SMPC since it needs the verification of the random number generator. Cryptoleq also depends on heuristic code-based obfuscation.
There is a Open RISC implementation based on idea of HEROIC~\cite{10.5220/0005955902390250}, but this is suffered from too much memory consumption because of big tables. The authors estimated it to be between hundreds of gigabytes to terabytes.

\subsection{Cingulata}
Cingulata~\cite{10.1145/2732516.2732520,Cingulata} is a framework to compile internal DSL in C++ into a Boolean circuit using High Level Synthesis (HLS)~\cite{10.1145/3243734.3243828,10.1145/2732516.2732520} and evaluate it over HE~\cite{10.1145/2732516.2732520,Cingulata}. 
HLS itself is not specialized for SMPC, but is mainly used for programming FPGA~\cite{10.1145/3356475}.
Because Cingulata requires to directly represent the function to be evaluated as a Boolean circuit, it neither support two-party SCO nor PF-SFE.

\subsection{Intel SGX}
Intel Software Guard Extensions (SGX) is a set of security-related instructions that are built into Intel CPUs. There are several secure computation frameworks built on this~\cite{10.1145/3133956.3134095,203255}. Intel SGX has almost no overhead for secure evaluation of the functions because all ciphertexts are decrypted in CPU and executed as plaintext program. However, its security relies on trust in Intel. Because there is no theoretical reason to trust in Intel when the Alice does not trust in cloud vendors like Amazon and Google. In other words, VSP and Intel SGX are based on different adversary models.
Also, there is recently reported vulnerability for Intel SGX~\cite{sgaxe}.

\subsection{Obfuscation}
Obfuscation aims to turn programs into ``unintelligible'' ones while preserving functionality to protect the function~\cite{10.1145/3234511}. There are two types of obfuscation, code-oriented one and model-oriented one~\cite{obfuscation,10.1145/2160158.2160159}. Code-oriented obfuscation converts the conventional program representation like C language to the obfuscated binary. The obfuscated binary is not encrypted, so it can be run directly on the physical processor. The difficulty of obtaining the evaluated function from the obfuscated binary is measured by non-cryptographic metrics~\cite{10.1145/268946.268962,10.5555/1362903.1362922} while the security of VSP and model-oriented obfuscation is measured by cryptographic one. In model-oriented obfuscation, the function to be evaluated is represented by the mathematical mode like Boolean Circuit, Turing machine, matrix branching program, etc. then evaluated by the cryptographical way like multilinear map~\cite{10.1007/978-3-642-40041-4_26}. Model-oriented obfuscation may include VSP and GarbledCPU in definition. However, we could not find the work which treats model-oriented obfuscation and GarbledCPU in the same paper.

\begin{table}[t]
\centering\caption{Resource requirements of CAHP-Ruby.}\label{tab:ruby-gate-counts}
\begin{tabular}{c|cccc|c}
\toprule
Gate   & IF & ID+WB & Ex  & Mem & Total \\
\midrule
AND    & 193 & 270 & 301 & 92  & 651  \\
ANDNOT & 59  & 110 & 44  & 0   & 223  \\
MUX    & 54  & 683 & 256 & 116 & 996  \\
NAND   & 300 & 336 & 356 & 10  & 1025 \\
NOR    & 4   & 77  & 31  & 0   & 90   \\
NOT    & 3   & 4   & 11  & 1   & 15   \\
OR     & 77  & 56  & 95  & 8   & 215  \\
ORNOT  & 39  & 142 & 40  & 8   & 195  \\
XNOR   & 3   & 9   & 40  & 0   & 51   \\
XOR    & 4   & 5   & 21  & 0   & 36   \\
\bottomrule
\end{tabular}
\end{table}

\begin{table}[t]
\centering\caption{Resource requirements of CAHP-Pearl}\label{tab:pearl-gate-counts}
\begin{tabular}{c|cccc|c}
\toprule
Gate   & IF & ID+WB & Ex  & Mem & Total \\
\midrule
AND    & 199 & 191 & 200 & 17 & 539 \\
ANDNOT & 48  & 69  & 37  & 0  & 126 \\
MUX    & 65  & 619 & 132 & 41 & 877 \\
NAND   & 309 & 313 & 365 & 10 & 937 \\
NOR    & 2   & 30  & 15  & 0  & 100 \\
NOT    & 3   & 3   & 11  & 0  & 22  \\
OR     & 83  & 9   & 104 & 8  & 148 \\
ORNOT  & 57  & 89  & 47  & 8  & 131 \\
XNOR   & 3   & 0   & 39  & 0  & 50  \\
XOR    & 4   & 0   & 22  & 0  & 23  \\
\bottomrule
\end{tabular}
\end{table}

\begin{table}[t]
\centering\caption{The number of CMUXs in CMUX Memory}\label{tab:cmux-gate-count}
\begin{tabular}{cc}
\toprule
Component      & \# of CMUXs   \\
\midrule
RAM Read Unit  & 4,080   \\
RAM Write Unit & 32,768 \\
ROM            & 8      \\
\midrule
Total          & 36,856  \\
\bottomrule
\end{tabular}
\end{table}

\section{Abstract Protocol Flow in two-party PF-SFE}
\label{sec:pf-sfe}

In this section, we explain how the protocol flow of VSP can be theoretically modified to do two-party PF-SFE.

{\bf Public/Private Data}: 
In this protocol, the function to be evaluated is provided by Bob and the input data is provided by Alice. The most important fact for understanding how VSP works in this protocol is that TFHE supports ``trivial'' ciphertexts. 
``Trivial'' here means the generation of them does not need any secret key nor random number generation. 
For example, trivial TLWE of 1 is $(\mathbf{0},\mu)$. In such a way, Bob can provide ROM and RAM data without the input of Alice.

\subsection{Abstract protocol workflow}
\begin{figure}[tb]
  \centering
  \includegraphics[width=0.7\linewidth]{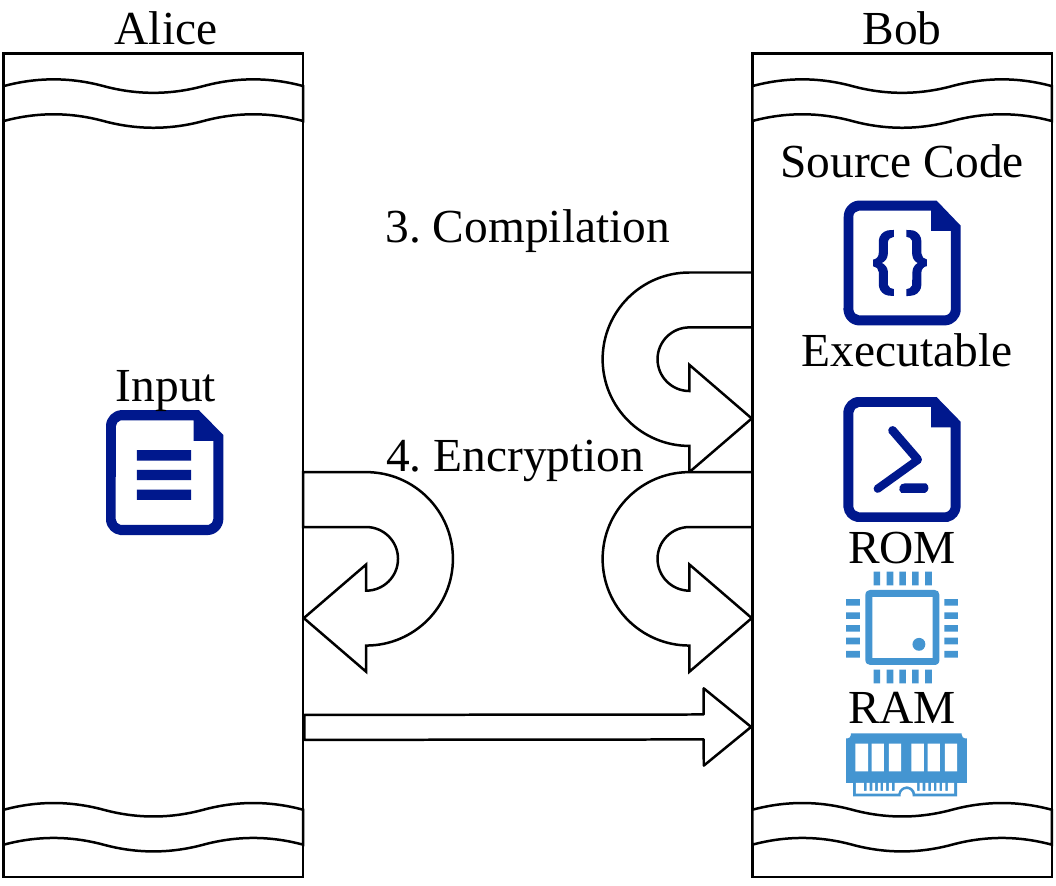}
  \caption{Protocol flow of PF-SFE}
  \label{fig:pfsfeworkflow}
\end{figure}

The visual image is shown in \cref{fig:pfsfeworkflow}. This shows only difference from two-party SCO case. Phase 1., 2. and 5. to 7. are the same as two-party SCO.

\begin{enumerate}
\setcounter{enumi}{2}
\item {\bf Compilation}: Bob compiles the source code of the desired function into the executable for the processor.
\item {\bf Encryption}: Alice encrypts the input and sends it to Bob. Bob encrypts the executable using trivial ciphertexts and combines it with the encrypted input to generate the encrypted ROM and RAM.
\end{enumerate}

\subsection{Security Analysis in Two-party PF-SFE}

Bob tries to reveal plaintexts of Bootstrapping Key, RAM, registers and each ciphertext of each wire, etc. ROM (the function to be evaluated) is not a target since it is provided by Bob. However, like two-party SCO case, this can be reduced to the hardness of decryption of ciphertexts of TFHE. Alice tries to reveal ROM, RAM, registers and ciphertexts of each wire, etc. Though they will not be provided to Alice,  Alice knows the result of the function, so if Bob uses always the same function and input, Alice can try to get the results for all possible inputs. Therefore, the protection of private information of Bob from Alice needs another method like indistinguishable obfuscation. This is beyond the scope of our proposed method.

We note that PF-SFE protocol does have the FHE malleability 
problem, since the program is provided by Bob. However, PF-SFE 
is still vulnerable to the termination problem mentioned in 
Section~4.1.

\section{Full Version of Evaluation}
\label{sec:fulleval}

The program Hamming takes two inputs: $a$ and $b$.
We used $a=10101010_{(16)}$ and $b=\mbox{deadbeef}_{(16)}$ in the evaluation.

We used two more programs except Hamming: Fibonacci and Brainf*ck.
Fibonacci takes a decimal digit $n$ as its command-line argument, and calculates $n$th Fibonacci number.
Here we used $n=5$.
Brainf*ck interprets code of brainf*ck, which is a esoteric programming language, and returns the result.
We inputted \verb|++++[>++++++++++<-]>++| to it, the result of which is 42. 

\subsubsection*{Gate count evaluation}

\cref{tab:ruby-gate-counts} and \cref{tab:pearl-gate-counts} show the gate requirements of
each stage of CAHP-Ruby and CAHP-Pearl.
Note that these values are calculated by synthesising the components separately.
Due to global optimizations in the synthesis software,
the numbers do not add up to the size of the entire processor circuit (the column ``Total'').

Also, \cref{tab:cmux-gate-count} shows the number of CMUXs in CMUX Memory components.

\subsubsection*{Memory consumption evaluation}

On p3.8xlarge, running our implementation consumes about 3.7 GB of main memory and about 0.6 GB per GPU. The most of memory consumption seems to be caused by holding Bootstrapping Key.

\subsubsection*{Runtime Performance Evaluation}
\cref{tab:full-program-evaluation} shows all the results of our runtime performance evaluation benchmarks.
\cref{tab:full-program-evaluation-1KiB} shows additional results in 1 KiB ROM and RAM setting.

\begin{table*}[tb]
  \caption{Runtime performance evaluation in full version. The more cycles a program takes, the less run-time it takes per cycle, because we can ignore the constant time it takes to initialize the system, for example, reset all registers.
  Also note that NaN means that the number of the trials is 1.}
  \label{tab:full-program-evaluation}
  \small\centering
  \begin{tabular}{ccccc|cccc}
  \toprule
Machine & \begin{tabular}{@{}c@{}}\# of\\V100\end{tabular} & Pipelining? & CMUX Memory? & Program & \begin{tabular}{@{}c@{}}\# of\\cycles\end{tabular} & Runtime~[s] & sec./cycle & \begin{tabular}{@{}c@{}}\# of\\tries\end{tabular}\\
\midrule

Sakura Koukaryoku & 1 & No & No & Fibonacci & 39 & $257.0 \pm 1.0$ & $6.590 \pm 0.027$ & 5 \\
Sakura Koukaryoku & 1 & No & No & Hamming & 936 & $5919.0 \pm 33.1$ & $6.324 \pm 0.035$ & 5 \\
Sakura Koukaryoku & 1 & No & No & Brainf*ck & 1905 & $12030.0 \pm 52.9$ & $6.313 \pm 0.028$ & 5 \\
Sakura Koukaryoku & 1 & No & Yes & Fibonacci & 39 & $101.9 \pm 0.7$ & $2.614 \pm 0.017$ & 5 \\
Sakura Koukaryoku & 1 & No & Yes & Hamming & 936 & $2232.1 \pm 1.7$ & $2.385 \pm 0.002$ & 5 \\
Sakura Koukaryoku & 1 & No & Yes & Brainf*ck & 1905 & $4537.0 \pm 5.7$ & $2.381 \pm 0.003$ & 5 \\
Sakura Koukaryoku & 1 & Yes & No & Fibonacci & 57 & $376.2 \pm 2.4$ & $6.600 \pm 0.042$ & 4 \\
Sakura Koukaryoku & 1 & Yes & No & Hamming & 1216 & $7809.0 \pm 45.8$ & $6.422 \pm 0.038$ & 4 \\
Sakura Koukaryoku & 1 & Yes & No & Brainf*ck & 2635 & $16870.0 \pm 87.7$ & $6.401 \pm 0.033$ & 4 \\
Sakura Koukaryoku & 1 & Yes & Yes & Fibonacci & 57 & $103.1 \pm 0.2$ & $1.810 \pm 0.004$ & 5 \\
Sakura Koukaryoku & 1 & Yes & Yes & Hamming & 1216 & $2045.0 \pm 4.6$ & $1.682 \pm 0.004$ & 5 \\
Sakura Koukaryoku & 1 & Yes & Yes & Brainf*ck & 2635 & $4431.0 \pm 10.1$ & $1.682 \pm 0.004$ & 5 \\
AWS p3.8xlarge & 2 & No & Yes & Fibonacci & 39 & $84.8 \pm NaN$ & $2.18 \pm NaN$ & 1 \\
AWS p3.8xlarge & 2 & No & Yes & Hamming & 936 & $1626.3 \pm NaN$ & $1.74 \pm NaN$ & 1 \\
AWS p3.8xlarge & 2 & Yes & Yes & Fibonacci & 57 & $73.9 \pm NaN$ & $1.30 \pm NaN$ & 1 \\
AWS p3.8xlarge & 2 & Yes & Yes & Hamming & 1216 & $1217.2 \pm NaN$ & $1.00 \pm NaN$ & 1 \\
AWS p3.8xlarge & 3 & Yes & Yes & Fibonacci & 57 & $71.0 \pm NaN$ & $1.25 \pm NaN$ & 1 \\
AWS p3.8xlarge & 3 & Yes & Yes & Hamming & 1216 & $1065.5 \pm NaN$ & $0.88 \pm NaN$ & 1 \\
AWS p3.8xlarge & 4 & No & Yes & Fibonacci & 39 & $86.7 \pm 0.3$ & $2.222 \pm 0.009$ & 3 \\
AWS p3.8xlarge & 4 & No & Yes & Hamming & 936 & $1455.7 \pm 0.3$ & $1.555 \pm 0.000$ & 3 \\
AWS p3.8xlarge & 4 & Yes & Yes & Fibonacci & 57 & $70.8 \pm 0.6$ & $1.242 \pm 0.010$ & 3 \\
AWS p3.8xlarge & 4 & Yes & Yes & Hamming & 1216 & $979.0 \pm 12.5$ & $0.805 \pm 0.010$ & 3 \\
AWS p3.2xlarge & 1 & No & Yes & Fibonacci & 39 & $105.9 \pm NaN$ & $2.72 \pm NaN$ & 1 \\
AWS p3.2xlarge & 1 & No & Yes & Hamming & 936 & $2330.7 \pm NaN$ & $2.49 \pm NaN$ & 1 \\
AWS p3.2xlarge & 1 & Yes & Yes & Fibonacci & 57 & $104.3 \pm NaN$ & $1.83 \pm NaN$ & 1 \\
AWS p3.2xlarge & 1 & Yes & Yes & Hamming & 1216 & $2068.0 \pm 6.0$ & $1.701 \pm 0.005$ & 2 \\
AWS p3.2xlarge & 1 & Yes & Yes & Brainf*ck & 2635 & $4464.5 \pm NaN$ & $1.69 \pm NaN$ & 1 \\
AWS p3.16xlarge & 4 & No & Yes & Fibonacci & 39 & $105.3 \pm NaN$ & $2.70 \pm NaN$ & 1 \\
AWS p3.16xlarge & 4 & No & Yes & Hamming & 936 & $1486.6 \pm NaN$ & $1.59 \pm NaN$ & 1 \\
AWS p3.16xlarge & 4 & Yes & Yes & Fibonacci & 57 & $89.0 \pm NaN$ & $1.56 \pm NaN$ & 1 \\
AWS p3.16xlarge & 4 & Yes & Yes & Hamming & 1216 & $1021.1 \pm NaN$ & $0.84 \pm NaN$ & 1 \\
AWS p3.16xlarge & 6 & No & Yes & Fibonacci & 39 & $118.2 \pm NaN$ & $3.03 \pm NaN$ & 1 \\
AWS p3.16xlarge & 6 & No & Yes & Hamming & 936 & $1449.4 \pm NaN$ & $1.55 \pm NaN$ & 1 \\
AWS p3.16xlarge & 6 & Yes & Yes & Fibonacci & 57 & $101.6 \pm NaN$ & $1.78 \pm NaN$ & 1 \\
AWS p3.16xlarge & 6 & Yes & Yes & Hamming & 1216 & $979.7 \pm NaN$ & $0.81 \pm NaN$ & 1 \\
AWS p3.16xlarge & 7 & Yes & Yes & Fibonacci & 57 & $110.6 \pm NaN$ & $1.94 \pm NaN$ & 1 \\
AWS p3.16xlarge & 7 & Yes & Yes & Hamming & 1216 & $974.2 \pm NaN$ & $0.80 \pm NaN$ & 1 \\
AWS p3.16xlarge & 8 & No & No & Fibonacci & 39 & $94.3 \pm 0.9$ & $2.417 \pm 0.023$ & 3 \\
AWS p3.16xlarge & 8 & No & No & Hamming & 936 & $1627.0 \pm 4.2$ & $1.739 \pm 0.004$ & 3 \\
AWS p3.16xlarge & 8 & No & Yes & Fibonacci & 39 & $133.1 \pm 0.9$ & $3.413 \pm 0.023$ & 3 \\
AWS p3.16xlarge & 8 & No & Yes & Hamming & 936 & $1440.0 \pm 2.5$ & $1.538 \pm 0.003$ & 3 \\
AWS p3.16xlarge & 8 & Yes & No & Fibonacci & 57 & $98.5 \pm 1.0$ & $1.728 \pm 0.017$ & 3 \\
AWS p3.16xlarge & 8 & Yes & No & Hamming & 1216 & $1566.0 \pm 9.7$ & $1.288 \pm 0.008$ & 3 \\
AWS p3.16xlarge & 8 & Yes & Yes & Fibonacci & 57 & $116.8 \pm 0.4$ & $2.050 \pm 0.007$ & 3 \\
AWS p3.16xlarge & 8 & Yes & Yes & Hamming & 1216 & $965.9 \pm 3.4$ & $0.794 \pm 0.003$ & 3 \\
AWS c5.metal & 0 & No & Yes & Fibonacci & 39 & $103.1 \pm 1.0$ & $2.642 \pm 0.025$ & 3 \\
AWS c5.metal & 0 & No & Yes & Hamming & 936 & $2342.0 \pm 13.3$ & $2.502 \pm 0.014$ & 3 \\
AWS c5.metal & 0 & Yes & Yes & Fibonacci & 57 & $131.4 \pm 0.0$ & $2.305 \pm 0.001$ & 3 \\
AWS c5.metal & 0 & Yes & Yes & Hamming & 1216 & $2773.0 \pm 2.8$ & $2.280 \pm 0.002$ & 3 \\

\bottomrule
  \end{tabular}
\end{table*}

\begin{table*}[tb]
  \caption{Runtime performance evaluation in 1 KiB ROM and RAM setting}
  \label{tab:full-program-evaluation-1KiB}
  \small\centering
  \begin{tabular}{ccccc|cccc}
  \toprule
Machine & \begin{tabular}{@{}c@{}}\# of\\V100\end{tabular} & Pipelining? & CMUX Memory? & Program & \begin{tabular}{@{}c@{}}\# of\\cycles\end{tabular} & Runtime~[s] & sec./cycle & \begin{tabular}{@{}c@{}}\# of\\tries\end{tabular}\\
\midrule
AWS p3.8xlarge & 4 & No & No & Fibonacci & 40 & $165.0 \pm 12.9$ & $4.124 \pm 0.323$ & 2 \\
AWS p3.8xlarge & 4 & No & No & Hamming & 937 & $3306.0 \pm 12.9$ & $3.528 \pm 0.014$ & 2 \\
AWS p3.8xlarge & 4 & No & Yes & Fibonacci & 40 & $114.6 \pm 0.6$ & $2.866 \pm 0.016$ & 3 \\
AWS p3.8xlarge & 4 & No & Yes & Hamming & 937 & $1733.4 \pm 1.1$ & $1.850 \pm 0.001$ & 3 \\
AWS p3.8xlarge & 4 & Yes & No & Fibonacci & 58 & $207.0 \pm 10.6$ & $3.575 \pm 0.183$ & 3 \\
AWS p3.8xlarge & 4 & Yes & No & Hamming & 1217 & $3804.0 \pm 18.6$ & $3.126 \pm 0.015$ & 3 \\
AWS p3.8xlarge & 4 & Yes & Yes & Fibonacci & 58 & $100.0 \pm 1.6$ & $1.726 \pm 0.027$ & 5 \\
AWS p3.8xlarge & 4 & Yes & Yes & Hamming & 1217 & $1217.0 \pm 2.4$ & $1.000 \pm 0.002$ & 5 \\
\midrule
AWS c5.metal & 0 & No & Yes & Fibonacci & 40 & $143.0 \pm 1.7$ & $3.582 \pm 0.042$ & 3 \\
AWS c5.metal & 0 & No & Yes & Hamming & 937 & $3230.0 \pm 40.6$ & $3.443 \pm 0.043$ & 3 \\
AWS c5.metal & 0 & Yes & Yes & Fibonacci & 58 & $190.0 \pm 3.1$ & $3.273 \pm 0.053$ & 3 \\
AWS c5.metal & 0 & Yes & Yes & Hamming & 1217 & $3940.0 \pm 68.4$ & $3.238 \pm 0.056$ & 3 \\
\bottomrule
  \end{tabular}
\end{table*}

\end{document}